\documentclass[11pt,aps,nofootinbib,prd,aps,epsf,floats,axodraw,amsmath,amssymb,amsfonts]{revtex4}
\usepackage{amsmath, amssymb}
\newcommand{\mathsym}[1]{{}}

\usepackage{graphicx}
\usepackage{amsmath}
\usepackage{amssymb}
\usepackage{bm}
\newcommand{\be}{\begin{equation}}
\newcommand{\ee}{\end{equation}}
\newcommand{\bea}{\begin{eqnarray}}
\newcommand{\eea}{\end{eqnarray}}

\newcommand{\rem}[1]{}
\newsavebox{\PSLASH}
 \sbox{\PSLASH}{$p$\hspace{-1.8mm}/}
 
\renewcommand{\theequation}{\thesection.\arabic{equation}}
\newcounter{saveeqn}
\newcommand{\add}{\addtocounter{equation}{1}}
\newcommand{\alpheqn}{\setcounter{saveeqn}{\value{equation}}%
\setcounter{equation}{0}%
\renewcommand{\theequation}{\mbox{\thesection.\arabic{saveeqn}{\alph{equation}}}}}
\newcommand{\reseteqn}{\setcounter{equation}{\value{saveeqn}}%
\renewcommand{\theequation}{\thesection.\arabic{equation}}}

 \newsavebox{\notrightarrow}
 \sbox{\notrightarrow}{$\to$\hspace{-4mm}/}
 
 \newsavebox{\PARTIALSLASH}
 \sbox{\PARTIALSLASH}{$\partial$\hspace{-1.6mm}/}
 
 \newsavebox{\ASLASH}
 \sbox{\ASLASH}{$A$\hspace{-2.1mm}/}
 
 \newsavebox{\KSLASH}
 \sbox{\KSLASH}{$k$\hspace{-1.8mm}/}
 
 \newsavebox{\LSLASH}
 \sbox{\LSLASH}{$\ell$\hspace{-1.8mm}/}
 
 \newsavebox{\QSLASH}
 \sbox{\QSLASH}{$q$\hspace{-1.8mm}/}
 
 \newsavebox{\DSLASH}
 \sbox{\DSLASH}{$D$\hspace{-2.2mm}/}
 
 \newsavebox{\DbfSLASH}
 \sbox{\DbfSLASH}{${\mathbf D}$\hspace{-2.8mm}/}
 
 \newsavebox{\DELVECRIGHT}
 \sbox{\DELVECRIGHT}{$\stackrel{\rightarrow}{\partial}$}
 
 \newcommand{\blue}{\IfColor{\textCadetBlue}{}}
\newcommand{\black}{\IfColor{\textBlack}{}}
\newcommand{\red}{\IfColor{\textRed}{}}
\newcommand{\green}{\IfColor{\textOliveGreen}{}}
\newcommand{\lila}{\IfColor{\textRedViolet}{}}

\begin{document}
\begin{flushright}
 [hep-th/gr-qc/math-ph]
\end{flushright}
\title{Brownian Motion in the Hilbert Space of Quantum States along with the Ricci Flow and Stochastically Emergent Einstein-Hilbert Action: Formulating a Well-Defined Feynman's Path-Integral Measure for Quantum Fields in the Presence of Gravity}

\author{Amir Abbass Varshovi}\email{ab.varshovi@sci.ui.ac.ir/amirabbassv@gmail.com/amirabbassv@ipm.ir}

\affiliation{Faculty of Mathematics and Statistics, Department of Applied Mathematics and Computer Science, University of Isfahan, Isfahan, IRAN.\\
   School of Mathematics, Institute for Research in Fundamental Sciences (IPM), P.O. Box: 19395-5746, Tehran, IRAN.\\
   Visitor Academic, Cyrus University, California, USA.}
\begin{abstract}
   \textbf{Abstract\textbf{:}} In this paper, we aim to interpret the background gravitational effects appearing in quantum field theory on curved space-time by studying the Brownian motion of quantum states along with the Hamilton-Perelman Ricci flow. It has been shown that the Wiener measure automatically contains the Einstein-Hilbert action and the path-integral formulation of the scalar quantum field theory on curved space-time at the first order of local approximations. This provides a well-defined formulation of the path-integral measure for quantum field theory in the presence of gravity. However, we establish that the emergence of Einstein-Hilbert action is independent of the matter field interactions and is a merely entropic/geometric effect stemming from the nature of the Ricci flow of the universe geometry. We also extract an explicit formula for the cosmological constant in terms of the Ricci flow and Hamilton's theorem for 3-manifolds. Then, we discuss the cosmological features of the FLRW solution in $\Lambda$CDM Model via the derived equations of the Ricci flow. We also argue the correlation between our formulations and the entropic aspects of gravity. Finally, we provide some theoretical evidence that proves the second law of thermodynamics is the basic source of gravity and probably a more fundamental concept. \\
       
\noindent \textbf{Keywords\textbf{:}} Wiener Stochastic Process, Fractal Norm, Laplace-Beltrami Operator, Closed Riemannian Manifold, Einstein-Hilbert Action, Hamilton's Theorem, Ricci Flow, 3-Manifold, Perelman's Entropy Functional, Entropic Force, Cosmological Constant, $\Lambda$CDM Model, The Second Law of Thermodynamics.
\end{abstract}

 \maketitle


\section{Introduction}
\label{introduction}

\par This paper tries to insert gravity into the measure-theoretic formulation of Feynman's path-integral measure we established recently in \cite{varshovi1}. To have a self-contained perusal this introduction is devoted to providing a brief description of the contents of \cite{varshovi1} and expressing the idea of its generalization to non-flat closed Riemannian manifolds. We remark that by employing the Wiener stochastic process \cite{wiener}\footnote{See \cite{folland} for a nice presentation of the Wiener stochastic process and its applications as a probability measure.} in the Hilbert space of quantum states on a flat manifold via the framework of fractality to renormalize the integration in infinite dimensions we have recently derived in \cite{varshovi1} a mathematically well-defined path-integral measure for scalar quantum field theory, the so-called \emph{Wiener fractal measure}. The mentioned consequence stands on a combined footing that consists of two basis:\\
\par \textbf{i)} The concept of \emph{fractality} of continuous functions (quantum states).
\par \textbf{ii)} The Wiener path-integral formulation of \emph{Brownian motion}.\\

\noindent Although the latter is a well-known topic in probability theory, the former needs more explanation. In fact, the fractality as the most pivotal concept in Wilsonian renormalization settling in the core of path-integral measure is an analytical property of continuous functions defined based on the Fourier-Laplace expansions revealing the weight of their self-similarity and non-differentiability \cite{varshovi1}. This property is, indeed, figured out by studying the analytic features of Weierstrass-like functions which we refer to as the \emph{fractal functions}.

\par The fractal functions are continuous but non-differentiable in their domain except on a small subset. We studied them according to the properties of the Weierstrass functions \cite{weierstrass} and their asymptotic behavior in Fourier coefficients. Following Hardy's results \cite{hardi} we obtained a non-linear differential equation for such asymptotic behaviors of fractal functions whose solution establishes that $f:S\to \mathbb R$ is a fractal function if for each $\sigma >0$, $\mu\geq 0$, and $\ell >0$ there exists some Fourier mode $n=(n_1,\cdots,n_D)\in \mathbb Z^D$, so that
\begin{equation} \label {1}
\int_0^{|f_n|}e^{\sigma (\mu^2+n^2)x^2/2 } dx > \ell,
\end{equation}

\noindent wherein $S=[-L,L]^D$ is a $D$-dimensional cube (or equivalently $D$-dimensional torus $S=\mathbb T^D$), $f_n$ is the $n$-th Fourier coefficient of $f$, and $n^2=\sum_{i=1}^D n_i^2$. Setting $\Delta=-\sum_{i=1}^D \partial_i^2$, we can readily rewrite (\ref {1}) in terms of the spectrum of the Laplacian operator. Let $\{\psi_{\lambda,i} \}_{\lambda \in \text{spec}\Delta}^{i\in D_\lambda}$ be the complete set of orthonormal eigenfunctions of $\Delta$ for eigenvalues $\lambda \geq 0$ and some possible degeneracy space $D_\lambda$ for each $\lambda$ labeled by $i$.\footnote{Thus, we have: $\int_S \psi_{\lambda,i}\psi_{\lambda',i'} d^Dx=\delta_{\lambda \lambda'}\delta_{ii'}$.} Then, based on (\ref {1}) $f:S\to \mathbb R$ will be a fractal function if for each $\sigma >0$, $\mu\geq 0$, and $\ell >0$ there exists some Fourier mode $(\lambda,i)$ so that
\begin{equation} \label {2'}
\int_0^{|f_{\lambda, i}|}e^{\sigma (\mu^2+\lambda)x^2/2 } dx > \ell,
\end{equation}

\noindent wherein $f_{\lambda,i}=\int_S \psi_{\lambda,i} f d^Dx$. In principle, the mechanism of which we employed in \cite{varshovi1} for working out the fractality condition (\ref {2'}) on the space manifold $S$ could be generalized to any compact Riemannian manifold $(M,g)$.\footnote{Along this paper, the Riemannian manifold $M$ is always assumed connected and orientable.} Actually, here we may concentrate on manifolds without boundary.\footnote{So the Neumann and Dirichlet boundary conditions for extracting the eigenvalues and the eigenspaces of the Laplace-Beltrami operator would be irrelevant.} It is well-known that if $(M,g)$ is a closed $D$-dimensional Riemannian manifold then the spectrum of the Laplacian $\Delta$ is discrete and consists of an increasing sequence $\{ \lambda_k\}_{k=1}^\infty$ of eigenvalues (with degeneracies) with $\lambda_1=0$ and $\lambda_k \to \infty$ as $k \to \infty$ \cite{yau}.

\par On the other hand, the eigenfunctions of the Laplacian operator on Riemannian manifold $(M,g)$ provide a pure point spectrum, which means that $L^2(M)=L^2(\Lambda^0M)$ admits a complete orthonormal basis consisting of eigenfunctions of $\Delta$, say $\{\psi_{\lambda,i} \}_{\lambda, i}$, with associated eigenvalues $\lambda$ and the label of degeneracy $i$.\footnote{The same assertion holds for $L^2(\Lambda^n M)$, $n>0$, and the $n$-eigenforms of $\Delta$.} This leads to a Fourier expansion of continuous functions\footnote{And also of continuous $n$-forms.} on compact Riemannian manifolds. An extensive amount of investigations have been made concerning these eigenvalues and their relationship to the background Riemannian geometry.\footnote{See for example \cite{weyl, weyl2, gallot, besse, donnelly, gilkey1, gilkey2, chavel, yau, li-yau, yau0, deshmukh, gilkey3, singer, berger, buser, cheeger, ledoux, lang, egidi, simon, henrot} and the references therein.} These studies pertain to upper and lower bounds for eigenvalues and their asymptotics which are of the most interest in our present study. There is a major conclusion that is valid for all compact Riemannian geometry; the so-called \emph{Weyl's asymptotic law} \cite{weyl, weyl2}, which asserts that $\lambda_k$ asymptotically behaves as $k^{2/D}$ as $k \to \infty$.

\par It is worth reminding that there are significant results for the first gap of the Laplacian eigenvalues:\\
\par \textbf{i)} Yau's theorem \cite{yau0, li-yau} which provides a lower bound for non-vanishing Laplacian eigenvalues on Riemannian manifolds with non-negative Ricci curvature as $\lambda_1 \geq \frac{\pi^2}{d^2}$, where $d$ is the diameter of $M$.
\par \textbf{ii)} Lichnerowicz's theorem \cite{gallot} for positive definite Ricci curvature, which says that $\lambda_1 \geq KD$ whenever $Ric_{\mu \nu} \geq (D-1)Kg_{\mu \nu}$.
\par \textbf{iii)} Simon's lower bound for an Einstein's manifold of $D\geq 3$ dimensions whose sectional curvature is bounded from below by a constant $K_0$, which states that $M$ is either isometric to $D$-sphere $S^D$ or else $\lambda_1 > 2DK_0$ \cite{simon}.\footnote{See \cite {deshmukh0, poor} for a more detailed review of the harmonic theory in Riemannian geometry and the lower bound of $\lambda_1$ in Einstein-like and constant scalar curvature manifolds.}
\par \textbf{iv)} And finally, Cheeger's isoperimetric constant puts an important geometric lower bound for $\lambda_1$, mostly referred to as \emph{Cheeger's inequality} \cite{cheeger}. This inequality also gives rise to an upper bound for $\lambda_1$ \cite{buser} which is essentially given in terms of the Ricci curvature \cite{ledoux}.\footnote{See \cite{lang} for the generalization of Cheeger's inequalities to the magnetic Laplacian $\Delta^\alpha$ defined for $d \mapsto d+\alpha$, $\alpha \in \Omega^1M$, on closed Riemannian manifold $M$. See also \cite{egidi} for more recent achievements about the eigenvalues of the magnetic Laplacian $\Delta^\alpha$.}\\

\par However, Weyl's asymptotic result guarantees that the asymptotic behavior of the Laplace-Beltrami operator on each $D$-dimensional closed Riemannian manifold equals that of flat $D$-torus $\mathbb T^D$, which we essentially considered in \cite {varshovi1}. Moreover, the fractality condition is in particular a local property that can be studied by means of the harmonic analysis on Riemannian manifolds. Indeed, according to Sobolev's theorem \cite{sobolev} any element of the Sobolev space $\mathcal {H}_t$, with $t = [D/2]+s$, for $0\leq s <1$, will uniformly converge to a continuous function whose derivative fails to converge everywhere on $M$,\footnote{See chapter 6 of \cite{warner} for detailed applications of Sobolev spaces on Riemannian manifolds.} hence it could be a Weierstrass-like function. Here, the Sobolev class of a function is studied locally on $M$ and the constant $s$ could be considered as the self-similarity scale of the fractal function, as was pointed out by Hardy \cite{hardi}.

\par Since the main criteria for studying the fractality in \cite{varshovi1} stem from the asymptotic behavior of the Fourier-Laplace coefficients\footnote{The Fourier expansion coefficients in terms of the eigenfunctions of the Laplacian operator on $S=[-L,L]^D \cong \mathbb T^D$.} of Weierstrass-like functions, we readily obtain similar differential equations for such coefficients on closed Riemannian manifold $M$. More precisely, we can restrict functions on local charts by employing the partition of unity on $M$, and write down the Laplacian operator $\Delta$ as a second-order differential equation whose the leading symbol coincides with the Laplacian of $\mathbb R^D$ or $\mathbb T^D$, as we considered in \cite{varshovi1}. Therefore, we can similarly deduce the generalized version of \emph{Theorem 2} of \cite{varshovi1} on closed Riemannian manifolds:\footnote{We omit the detailed arguments of the issue here and refer the reader to sections II and III of \cite{varshovi1} for a more comprehensive discussion.}\\

\par \textbf{Theorem 1;} \emph{Let $(M,g)$ be a closed Riemannian manifold with Laplacian operator $\Delta=dd^\dag+d^\dag d$. Let $\{\psi_{\lambda,i} \}_{\lambda \in \text{spec}\Delta}^{i \in D_{\lambda}}$ be the complete set of orthonormal eigenfunctions of $\Delta$ for eigenvalues $\lambda \geq 0$ and some possible degeneracy spaces $D_\lambda$ labeled by $i$. That is, $\int_M \psi_{\lambda,i}\psi_{\lambda',i'} d\Omega_g=\delta_{\lambda \lambda'} \delta_{ii'}$, for $d\Omega_g$ the Riemannian volume form on $M$. Then, $f:M\to \mathbb R$ is a fractal function if for each $\sigma >0$, $\mu\geq 0$, and $\ell >0$ there exists some Laplacian Fourier mode $(\lambda,i)$ so that}
\begin{equation} \label {2}
\int_0^{|f_{\lambda, i}|}e^{\sigma (\mu^2+\lambda)x^2/2 } dx > \ell,
\end{equation}

\noindent \emph{wherein $f_{\lambda,i}=\int_M f \psi_{\lambda,i} d\Omega_g$ is the corresponding Fourier-Laplace coefficient.}\\

\par Let $\mathbf C$ be the set of the continuous functions on $M$ that admit Fourier-Laplace expansion, i.e., $\mathbf C$ consists of functions that satisfy the following equality almost everywhere on $M$;\footnote{As we assume for $L^2(M)$, we consider two functions to define the same element of $\mathbf C$ when they are equal almost everywhere on $M$.}
\begin{equation} \label {ezafe-1}
f=\sum_{\lambda, i} f_{\lambda, i}  \psi_{\lambda,i}.
\end{equation}

\noindent Obviously, we have $ C^1(M) \subset \mathbf C $, but, however, there are Sobolev's functions in $\mathcal {H}_t$, with $t = [D/2]+s$, $0\leq s <1$, that fall outside $C^1(M)$. Such elements are, in fact, fractal or Weierstrass-like functions. The fractal metric $d_{\mu,\sigma}$ with the definition
\begin{equation} \label {ezafe-2}
~~~~~~~~~~~~~~~~~~~~~~~d_{\mu,\sigma}(f,g)=\text{Sup}_{\lambda, i} \Big|  \int_{g_{\lambda,i}}^{f_{\lambda,i}} e^{\sigma(\mu^2 + \lambda)x^2/2} dx \Big|,~~~~~~~~~~f,g \in \mathbf C
\end{equation}

\noindent turns $({\mathbf C},d_{\mu,\sigma})$ into a metric space whose topology puts the fractal functions in far distant regions of ${\mathbf C}$ \cite{varshovi1}. Actually, we can easily see that:\\

\par \textbf{a)} $d_{\mu,\sigma}(f,g)<\infty$ if $f$ and $g$ are both non-fractal;
\par \textbf{b)} $d_{\mu,\sigma}(f,g)=\infty$ if only one of the two functions $f$ and $g$ is fractal.\\

\noindent The fractal metric $d_{\mu,\sigma}$ provides a quantitative criterion for evaluating the fractality of continuous functions in $\mathbf C$. In fact, if $f_0 \in \mathbf C$ is the zero function, then the fractality of $f \in \mathbf C$ is defined as
\begin{equation} \label {amir}
\text{Fractality~of~} f:=d_{\mu,\sigma}(f,f_0).
\end{equation}

\noindent In principle, if we define the \emph{(massive) fractal norm} $\ell_{\sigma,\mu}$ on the phase space of $\mathbf C$ as
\begin{equation} \label {ezafe-3}
\ell_{\sigma,\mu}(f_{\lambda,i}) =\Big| \int_0^{f_{\lambda,i}}e^{\sigma(\mu^2 + \lambda)x^2/2}dx \Big|,
\end{equation}

\noindent and replace the Lebesgue measure of the phase space with the product measure induced by the fractal norm $\ell_{\sigma,\mu}$, then we could easily apply the fractality (\ref {amir}) in studying the stochastic properties of continuous functions in $\mathbf C$. Actually, this would happen when we let $\ell_{\sigma,\mu}(f_{\lambda,i})$ grow up as far as possible. This idea is the most technical approach to Wilsonian renormalization in the path-integral formulation and is the main mechanism that we followed in \cite{varshovi1} to work out a mathematically well-defined reformulation of Feynman's path-integral measure for formulating the theory of scalar quantum fields. In this way, our main motivation for defining the metric space $({\mathbf C},d_{\mu,\sigma})$ and the fractal norm $\ell_{\sigma,\mu}$ becomes clearer. In other words, as we have shown in \cite{varshovi1}, if $M$ is flat, then the Brownian motion in the Hilbert space of quantum states on $M$ will provide a mathematically well-defined path-integral measure for Feynman's path-integral of scalar quantum fields provided the measure of $\mathbf C$ is induced by the fractal norm $\ell_{\sigma,\mu}$. Here, $L^2(M)$, the standard Hilbert space of the quantum states on $M$, is indeed the closure of $\mathbf C$ within the $L^2$ topology.

\par Employing \textbf{Theorem 1} we can accomplish the Wiener stochastic process for the fractality of the quantum states (as continuous functions) on the general closed Riemannian manifold $(M,g)$. Through the next section, we follow the strategy of \cite{varshovi1} on a given closed Riemannian manifold (as the manifold of space) and extract the generalized result on curved spaces. Actually, we will establish that an immediate generalization of the Wiener Brownian process for propagating the continuous functions (quantum states) on $M$ along the time direction will lead to Feynman's path-integral formulation of scalar quantum field theory on a static curved space-time. Thus, considering the fractality via the Wilsonian RG flow within the functional measure of the Wiener path-integral will result in Feynman's path-integral formulation of scalar quantum field theory with a static background gravity.

\par However, the main objective of this paper is accomplished in section III. In this section, the Wiener fractal measure is implemented for dynamical space-time with some time-dependent metric $g_{ij}(t)$. We will establish that the Einstein-Hilbert action can be rigorously worked out from an entropic evolution of the space geometry due to the Hamilton-Perelman Ricci flow. Moreover, we will establish that the emergence of gravity in this path-integral formulation is independent of the quantum field theory and is merely an entropic effect. Therefore, although the expectation value of the stress-energy tensor of the involved quantum matter fields will affect the metric of space $M$ via Einstein's field equation, the time evolution of the metric is imposed by an entropic law that stems from the second law of thermodynamics. Hence, the Ricci flow could be assumed as a natural gauge fixing term for Einstein's field equation.

\par Actually, the Brownian motion is intrinsically an entropic effect that is subject to the second law of thermodynamics. Hence, if we consider the Brownian motion of quantum states, we must be concerned about an underlying entropic force that causes the evolution of the background geometry. This entropic evolution of the geometry of manifolds is best understood via the Hamilton-Perelman Ricci flow of the Riemannian metric \cite{hamilton1, perelman}. Thus, the Wiener stochastic process of fractality must address the Ricci flow of the metric of the Riemannian manifold $M$. Consequently, we will see in section III that:\\

\par \emph{To study the Brownian motion of the quantum states on a closed Riemannian manifold, one has to consider the Ricci flow of the underlying Riemannian geometry within the Wiener fractal measure, and this consideration will result in the appearance of Einstein-Hilbert action in the Feynman's path-integral measure for formulating scalar quantum fields on the curved space-time manifold. Therefore, the underlying gravitational effects are the most natural expected background phenomena when one aims to study the Brownian motion of quantum states on an entropic evolving curved closed Riemannian manifold}.\\

\par The entropic description of gravitational effects studied in section III leads to a semi-classical theory of gravity, which we refer to as the \emph{Wiener fractal gravity}. In section IV we study the cosmological results of this extracted stochastic-entropic theory of gravity and extract an exact entropic/geometric formula for the cosmological constant via Hamilton's theorem for 3-manifolds \cite{hamilton1}. Next, we will give an entropic interpretation of inflation and the dark energy. Comparison with the $\Lambda$CDM Model and the admitted FLRW solution in cosmology is concerned due in this section. Finally, in section V we will study the entropic features of the Ricci flow according to Perelman's seminal paper \cite{perelman}, and will conclude that the second law of thermodynamics could be regarded as the basic source of gravitational effects in nature. We then compare the Wiener fractal gravity with Verlinde's entropic gravity, stochastic gravity and Horava-Lifshitz gravity. 

\par
\section{Wiener Fractal Measure and Quantum Fields on Curved Space-time}
\setcounter{equation}{0}

\par As mentioned above, \textbf{Theorem 1} lets us define the (massive) fractal norm on Riemannian manifold $(M,g)$ as;
\begin{equation} \label {3}
\ell_{\sigma,\mu}(f_{\lambda,i}) =\int_0^{|f_{\lambda,i}|}e^{\sigma(\mu^2 + \lambda)x^2/2}dx.
\end{equation}

\par Let $\mathcal N \in \mathbb N$ and define $\mathbf C_{\mathcal N}$ to be the collection of real functions $f:M \to \mathbb R$ in $\mathbf C$ which are finite linear combinations of Laplacian eigenfunctions $\psi_{\lambda,i}$ with eigenvalues $0\leq \lambda \leq \mathcal N$. Each individual element of $\mathbf C_{\mathcal N}$ is conventionally called an \emph{$\mathcal N$-bounded function (on $M$)}. Then, the fractal norm (\ref {3}) leads to a \emph{Lebesgue fractal measure} $d\mu_{LF}$ on $\mathbf C_{\mathcal N}$ as:
\begin{equation} \label {4}
d\mu_{LF}=\prod^{D_\mathcal N} dx_{\lambda,i},~~~~~~~~~~dx_{\lambda,i}=e^{\sigma(\mu^2+\lambda)f_{\lambda,i}^2/2}df_{\lambda,i},
\end{equation} 

\noindent wherein $D_\mathcal N$ is the dimension of $\mathbf C_{\mathcal N}$. Since $M$ is compact $\mathbf C_{\mathcal N}$ is finite-dimensional (i.e. $D_\mathcal N<\infty$), hence $d\mu_{LF}$ is well-defined. Now, we are ready to study the Brownian motion of $\mathcal N$-bounded functions by means of the Wiener stochastic process and the background Lebesgue fractal measure $d\mu_{LF}$. This Brownian motion is defined by means of an appropriate Wiener measure on $\mathbf C_{\mathcal N}$ for Riemannian manifold $(M,g)$. Actually, the definition is straightforward and is given upon what we have obtained in \cite{varshovi1}. We remember that the symmetric Wiener measure for a stochastic process in target space $\mathbb R^D$ from the initial point $x_I$ at $t_0=-T$ to the final destination $x_F$ at $t_{N+1}=T$, within the time slicing $-T=t_0<t_1< \cdots <t_N<t_{N+1}=T$ is:
\begin{equation} \label {5---}
\begin{gathered}
dW(t_N,\cdots,t_1)=\left(4\pi T\right)^{D_{\mathcal N}/2}~\exp\left( |x_F-x_I|^2/4T\right)\\
\times \frac{1}{\left( 2\pi \tau_{N+1}\right)^{D_{\mathcal N}/2}} e^{-|x_F-x_N|^2/2\tau_{N+1}} d^Dx_N \frac{1}{\left( 2\pi \tau_N\right)^{D_{\mathcal N}/2}} e^{-|x_N-x_{N-1}|^2/2\tau_N}\times \cdots\\
\times \cdots \times \frac{1}{\left( 2\pi \tau_2\right)^{D_{\mathcal N}/2}} e^{-|x_2-x_1|^2/2\tau_2}d^Dx_1 \frac{1}{\left( 2\pi \tau_1 \right)^{D_{\mathcal N}/2} } e^{-|x_1-x_I|^2/2\tau_1},~~~~~~~~~~~~~~
\end{gathered}
\end{equation}

\noindent for $x_i=x(t_i)$, $\tau_i=t_i-t_{i-1}$, and $|x_i-x_{i-1}|^2=\sum_{j=1}^D (x^j_i-x^j_{i-1})^2$. Therefore, the symmetric Wiener measure for the Brownian motion of the fractality of continuous functions on $M$ restricted to $\mathbf C_{\mathcal N}$ would be (assuming $x_I=x_F=0$):
\begin{equation} \label {5}
\begin{gathered}
dW(t_N,\cdots,t_1)=\left( 4\pi T \right)^{D_{\mathcal N}/2} \left( \prod_{k=1}^{N+1} \Xi_k \right) \times \left( \prod_{k=1}^N d^{D_{\mathcal N}} f_{\lambda,i}(t_k) \right),
\end{gathered}
\end{equation}

\noindent for 
\begin{equation} \label {6}
\begin{gathered}
\Xi_k=\frac{1}{\left( 2\pi \tau_{k}\right)^{D_{\mathcal N}/2}}
\times \exp \left( -\frac{\tau_{k}}{2}\sum_{ \lambda \in \text{spec}\Delta ~ , ~\lambda \leq \mathcal N} ~~ \sum_{i \in D_\lambda}\left(\frac{1}{ \tau_k } \int_{f_{\lambda,i}(t_{k-1})}^{f_{\lambda,i}(t_k)} e^{\sigma (\lambda + \mu^2)x^2 /2} dx \right)^2 \right)  \\
~~~~~~~~~~~\times \exp\left( \frac{\sigma}{2} \int_M \left( g^{ij}\partial_i f(t_{k-1}) \partial_j f(t_{k-1})+ \mu^2 f^2(t_{k-1}) \right) d\Omega_g \right).
\end{gathered}
\end{equation}

\noindent We refer to (\ref {5}) as the \emph{Wiener fractal measure}. Being of a non-local formula, this probability measure could be turned into a local structure by employing the following approximation as $N \to \infty$:\\

\par \textbf{Strategy A:} \emph{When $\tau_k \to 0$ (i.e. $N \to \infty$) we can simply apply the following replacement within the Wiener fractal measure:}
\begin{equation} \label {7}
\frac{1}{ \tau_k } \int_{f_{\lambda,i}(t_{k-1})}^{f_{\lambda,i}(t_k)} e^{\sigma (\lambda + \mu^2)x^2 /2} dx ~~~~~~~~~~\mapsto ~~~~~~~~~~\partial_t f_{\lambda,i} (t_{k-1}),
\end{equation}

\noindent \emph{which leads to:}
\begin{equation} \label {8}
\begin{gathered}
\Xi_k=\frac{1}{\left( 2\pi \tau_{k} \right)^{D_{\mathcal N}/2} }
 \exp \left( -\tau_k \int_M \frac{1}{2}\Big\{ \left(\partial_t f (t_{k-1}) \right)^2 -\frac{\sigma}{\tau_k} \left( g^{ij}\partial_i f(t_{k-1}) \partial_j f(t_{k-1})+ \mu^2 f^2(t_{k-1}) \right) \Big\} d\Omega_g\right).
\end{gathered}
\end{equation}
~
\par Actually, upon the strategy \textbf{A} we readily find:
\begin{equation} \label {9}
\begin{gathered}
\prod_{k=1}^{N+1} \Xi_k= \left( \frac{1}{\left( 2\pi \theta \right)^{D_{\mathcal N}/2} } \right)^{N+1}
 \exp \left( - \int_{\mathcal M} \frac{1}{2}\Big\{ \left(\partial_t f \right)^2 -\frac{\sigma}{\theta} \left( g^{ij}\partial_i f \partial_j f+ \mu^2 f^2 \right) \Big\} dt \wedge d\Omega_g \right),
\end{gathered}
\end{equation}

\noindent wherein $\mathcal M=[-T,T] \times M$ is the space-time continuum,\footnote{Indeed, as we pointed out in \cite{varshovi1} we distinguish between terminologies \emph{spacetime} and \emph{space-time}, for which the former is realized to admit space and time dimensions substantially as an identical essence, whereas the latter embraces an inherent difference between them.} and $\theta =dt=2T/N$ ($\to 0$) is the unit of the uniform time slicing. Set $c=\sqrt{\sigma/\theta}$ to be the speed of light in vacuum, then we obtain:
\begin{equation} \label {9'}
\begin{gathered}
\prod_{k=1}^{N+1} \Xi_k= \left( \frac{1}{\left( 2\pi \theta \right)^{D_{\mathcal N}/2} } \right)^{N+1}
 \exp \left( - c \int_{\mathcal M} \frac{1}{2}\Big\{ \mathbf g^{\mu \nu }\partial_\mu f \partial_\nu f - \mu^2 f^2 \Big\} d\Omega \right),
\end{gathered}
\end{equation}

\noindent where $d\Omega=dx^0\wedge d\Omega_g$ is the Riemannian volume form on the space-time continuum $\mathcal M$ for the induced time-independent metric
$$\mathbf g=\mathbf g_{\mu \nu} dx^\mu \otimes dx^\nu=dx^0 \otimes dx^0 -g_{ij} dx^i \otimes dx^j$$
\noindent with $x^0=c t$. Note that $\mathbf g$ is a Lorentzian metric with signature $(+,-,-,-)$, hence the theory is locally Lorentz invariant. If one defines $f=\gamma^{-1} \phi$ for normalization factor $\gamma=\sqrt {\hbar c}$ we read:
\begin{equation} \label {9''}
\begin{gathered}
\prod_{k=1}^{N+1} \Xi_k= \left( \frac{1}{\left( 2\pi \theta \right)^{D_{\mathcal N}/2} } \right)^{N+1}
 \exp \left( - \frac{1}{\hbar} \int_{\mathcal M} \frac{1}{2}\Big\{ \mathbf g^{\mu \nu }\partial_\mu \phi \partial_\nu \phi - \frac{m^2c^2}{\hbar^2} \phi^2 \Big\} d\Omega \right),
\end{gathered}
\end{equation}

\noindent for $m=\hbar \mu/c$ to be the physical mass.\footnote{For more detailed discussion about the relations between fractal parameters $\sigma$ and $\mu^2$ and the physical constants $\hbar$ and $c$ see \cite{varshovi1}.} Hence, for the Feynman measure
\begin{equation} \label {10}
\frak D \phi=\left( 4\pi T \right)^{D_{\mathcal N}/2} \left( \frac{1}{\left( 2\pi \theta \right)^{D_{\mathcal N}/2} } \right)^{N+1} \gamma^{-ND_{\mathcal N}} \left( \prod_{k=1}^N d^{D_{\mathcal N}} \phi_{\lambda,i}(t_k) \right),
\end{equation}

\noindent the Wiener fractal measure with strategy \textbf{A} becomes:
\begin{equation} \label {11}
dW_{\textbf{A}}(t_N,\cdots,t_1)=  \exp \left( - \frac{1}{\hbar} \int_{\mathcal M} \frac{1}{2}\Big\{ \mathbf g^{\mu \nu }\partial_\mu \phi \partial_\nu \phi - \frac{m^2c^2}{\hbar^2} \phi^2 \Big\} d\Omega \right) \frak D \phi.
\end{equation}

\par However, as we discussed in \cite{varshovi1} the replacement (\ref {7}) causes the heat kernel of the Gaussian term of the Wiener measure to be replaced with a quadratic function which is not always dominating. Therefore, the Wiener fractal measure (\ref {11}) is not well-defined over $\mathbf C_{\mathcal N}$. To handle this subtlety, we should transfer to complex analysis. As we pointed out in \cite{varshovi1} incorporating an infinitesimal heat kernel term after analytic continuation of the Wiener fractal measure would compensate some of the pathologies emerging from the replacement (\ref {7}). This complexification is conducted via the following equality:
\begin{equation} \label {12}
\lim_{\varepsilon \to 0} \Big| \int_{\mathbb R} e^{(ia-\varepsilon)x^2} dx\Big|=\int_{\mathbb R} e^{-|a|x^2} dx.
\end{equation}

\noindent Then, the next strategy is:\\

\par \textbf{Strategy B:} \emph{To compensate for the pathology of strategy \textbf{A} we should employ the complex version of the Wiener fractal measure with an augmented heat kernel $i\varepsilon$-term due to (\ref {12}). Hence, we must obtain:}
\begin{equation} \label {13}
\begin{gathered}
dW(t_N,\cdots,t_1)\approx dW_{\textbf{A,B}}(t_N,\cdots,t_1) \\
= \exp \left(  \frac{i}{\hbar} \int_{\mathcal M} \frac{1}{2}\Big\{ \mathbf g^{\mu \nu }\partial_\mu \phi \partial_\nu \phi - \frac{(m^2-i \varepsilon) c^2}{\hbar^2} \phi^2 \Big\} d\Omega \right) \frak D \phi,\\
= \exp \left(  \frac{i}{\hbar} \int_{\mathcal M} \frac{1}{2} \sqrt{|\det \mathbf g|} \Big\{ \mathbf g^{\mu \nu }\partial_\mu \phi \partial_\nu \phi - \frac{(m^2-i \varepsilon) c^2}{\hbar^2} \phi^2 \Big\} d^{D+1}x \right) \frak D \phi,
\end{gathered}
\end{equation}

\noindent \emph{within the assumption of $\varepsilon \to 0$.}\\

\par In particular, after employing strategies \textbf{A} and \textbf{B} we have effectively worked out the path-integral formulation of the scalar quantum fields on a static curved space-time. However, we must insist that the only mathematically well-defined path-integral measure that thoroughly describes the Brownian motion of quantum states on the Riemannian manifold $M$ is the probability measure of (\ref{5}). Hence, $dW_{\textbf{A,B}}$, although is a finite measure, lacks the effect of non-local terms we omitted in strategy \textbf{A} and gives a partly true explanation of the Brownian motion in the Hilbert space of quantum states up to the first local approximation.

\par We should remind that a well-defined (i.e. renormalizable) quantum field theory on the static curved space-time $\mathcal M$ is defined for a sequence of Wiener integrable functionals, say $F^{(\mathcal N)}(t_N, \cdots, t_1)=\prod_{k=1}^N\exp\left( \tau_k \int_{M} V^{(\mathcal N)}(\phi(t_{k-1})) d\Omega  \right)$, usually referred to as the interaction terms with (local) potentials $V^{(\mathcal N)}(\phi(t))$, each of which is defined on $\mathbf C_{\mathcal N}$, so that the Wiener path-integral
\begin{equation} \label {14}
\begin{gathered}
I_{\mathcal N}=\int_{\mathbf C_{\mathcal N}} F^{(\mathcal N)}(t_N, \cdots, t_1) \big\{\phi(t_{i_1})\cdots \phi(t_{i_n}) \big\} dW(t_N,\cdots,t_1)\\
~~~~~~~\approx \int_{\mathbf C_{\mathcal N}} F^{(\mathcal N)}_{\textbf{B}}(t_N, \cdots, t_1) \big\{\phi(t_{i_1})\cdots \phi(t_{i_n}) \big\} dW_{\textbf{A,B}}(t_N,\cdots,t_1)
\end{gathered}
\end{equation}

\noindent results in a finite amplitude as $\mathcal N \to \infty$ (while $N$ grows up as an appropriate monotone increasing function of $\mathcal N$) for each $\{i_k\}_{k=1}^n$, wherein $F_{\textbf{B}}^{(\mathcal N)}(t_N, \cdots, t_1)=\prod_{k=1}^N\exp\left( -i\tau_k \int_{M} V^{(\mathcal N)}(\phi(t_{k-1})) d\Omega  \right)$. As we discussed in \cite{varshovi1} negative renormalizable potential terms $V^{(\mathcal N)}$ always lead to finite and predictable expectation values $I_{\mathcal N}$ as $N (\mathcal N) \to \infty$. Indeed, the negativity ensures that the energy levels are bounded from below, and the renormalizability puts forward a reasonable correlation between $V^{(\mathcal N)}$s (and the corresponding coupling constants) at different regimes of energy due to $\mathcal N$.\footnote{For more details about the renormalizability of a scalar quantum field theory within the Wiener fractal mechanism see \cite{varshovi1}. Also see \cite{holland, dewitt, wald1, wald2} for more discussions about quantization, regularization, renormalization, Casimir effect, Bogoliubov transformation, and the $S$-matrix components of such theories in the presence of classical gravity. Moreover, we refer the readers to \cite{hu} for an interesting review on the topic.}

\par In particular, a similar approach to the above mechanism could be employed to work out the Maxwell and the Yang-Mills actions on curved space-time $\mathcal M$ \cite{varshovi2}. Actually, for such gauge theories, one should replace the Fourier expansion of the Laplacian operator with that of the generalized Bochner Laplacian defined on principal bundles over the space manifold $M$. Therefore, one may claim that the Standard Model, as the admitted prescription for the fundamental forces of nature, including gauge fields, fermions, and the scalar Higgs boson, could be described and interpreted fascinatingly as a Brownian process of geometric structures over the space manifold (i.e. the universe). In the next section, we will show that the gravitation effect, within the framework of general relativity, is spectacularly explained by evolving the cosmos geometry via an entropic equation.

\par
\section{Einstein-Hilbert Action and the Ricci Flow of the Space Metric}
\setcounter{equation}{0}

\par Let us revisit the symmetric Wiener measure (\ref {5---}) once again. Suppose that we need to perform a linear transformation $F_i:\mathbb R^D \to \mathbb R^D$ at each intermediate time section $t_i$, $1\leq i \leq N$. Then, we have:
\begin{equation} \label {b-1}
\begin{gathered}
dW(t_N,\cdots,t_1)=\frak C_N  \frac{1}{\left( 2\pi \tau_N \right)^{D/2}} e^{-|\Delta x'_N|^2/2\tau_N} d^D\Delta x'_N\times
 \cdots \times \frac{1}{\left( 2\pi \tau_1 \right)^{D/2} }e^{-|\Delta x'_1|^2/2\tau_1} d^D\Delta x'_1\\
 =\frak C_N  \frac{1}{\left( 2\pi \tau_N \right)^{D/2}} e^{-|F_N(\Delta x_N)|^2/2\tau_N} d^DF_N(\Delta x_N)\times
 \cdots \times \frac{1}{\left( 2\pi \tau_1 \right)^{D/2}}e^{-|F_1(\Delta x_1)|^2/2\tau_1} d^DF_1(\Delta x_1)\\
 =\frak C_N  \frac{1}{\left( 2\pi \tau_N \right)^{D/2} } e^{-|F_N(\Delta x_N)|^2/2\tau_N} J_N d^D\Delta x_N\times
 \cdots \times \frac{1}{\left( 2\pi \tau_1 \right)^{D/2} }e^{-|F_1(\Delta x_1)|^2/2\tau_1} J_1 d^D\Delta x_1,~~~~
\end{gathered}
\end{equation}

\noindent for $\Delta x_i=x_i-x_{i-1}$, and $\frak C_N=\left(\frac{2T}{\tau_{N+1}}\right)^{D/2}~\exp\left( |x'_F-x'_I|^2/4T-|x'_F-x'_N|^2/2\tau_{N+1}\right)$,\footnote{The contribution of $x_N$ in $\frak C_N$ will ingage it in the integration on the last time section, however, as $N \to \infty$, $x'_N$ would be actually equal to $x'_F$. In fact, at the first approximation, $|x'_F-x'_N|^2/2\tau_{N+1}$ can be easily removed from $\frak C_N$ in the calculations. Consequently, $\frak C_N$ is considered as a constant in the Wiener measure (\ref {b-1}).} where $J_i$ is the Jacobian determinant of transformation $F_i$, and $\Delta x'_i=F_i(\Delta x_i)$. Here, we have assumed that $\{x'_i\}$ and $\{x_i\}$ are respectively the static and the dynamic coordinate systems on $\mathbb R^D$.

\par We can simply employ (\ref {b-1}) for a dynamical geometry of the space manifold $M$. Indeed, this geometric evolution is assumed to be absolutely independent of the topology of $M$, but rather it would have been related to the differential geometric structures, such as the Riemannian metric $g$.\footnote{To see the corresponding topological dependence see \cite{top1, top2, top3}.} Therefore, since the Wiener fractal measure was worked out for the Fourier expansion in terms of the Laplacian eigenfunctions on $(M,g)$, the time evolution of $g$ would cause some change of basis within the measure as we explained and established in (\ref {b-1}). The next theorem provides a consistent framework for such transformations versus infinitesimal variation of the metric.\footnote{For the variation of eigenvalues and eigenfunctions of the Laplacian (and its generalized versions such as $(p,q)$-Laplacian and the Witten Laplacian) versus the metric evolution see \cite{azami, besse, berger2, hwang, dicerbo, berger et al, lic, hou, fang, cao, elsoufi} and the references therein.}\\

\par \textbf{Theorem 2 (ref. \cite{berger2});} \emph{Let $g(t) = g+t h$, $|t| \ll 1$, be a small variation of the metric on Riemannian manifold $(M,g)$, for $h$ a symmetric $(0,2)$-tensor field on $M$. Set $\Delta^{g(t)}$ and $d\Omega_{g(t)}$ to be respectively the Laplacian operator and the Riemann volume form due to the metric $g(t)$. Let $\lambda$ be an eigenvalue of $\Delta=\Delta^{g(0)}$ with degeneracy space $D_\lambda$ labeled by $i$. Then, there exists a unique $\psi_{\lambda,i}(t) \in C^\infty(M)$ for each $\lambda \in \emph{spec} \Delta$ and $i\in D_\lambda$ such that;}
\par \textbf{a)}  \emph{$\Delta^{g(t)}\psi_{\lambda, i}(t)= \lambda_i(t)\psi_{\lambda,i}(t)$, $\lambda_i(t) \geq 0$.} 
\par \textbf{b)} \emph{$\lambda_i(t)$ and $\psi_{\lambda,i}(t)$ depend analytically on $t$ with $\lambda_i(0) = \lambda$ for each $i \in D_\lambda$.}
\par \textbf{c)} \emph{$\{\psi_{\lambda,i}(t)\}^{i \in D_\lambda}_{\lambda \in \emph{spec} \Delta}$ is an orthonormal basis with respect to integration on $M$ with $d\Omega_{g(t)}$.}\\

\par As we explained above, the Wiener Brownian motion is actually an entropic force whose origin stems from the second law of thermodynamics. On the other hand, according to Perelman \cite{perelman} the second law of thermodynamics causes the geometric evolution of the underlying manifold via the Hamilton-Perelman Ricci flow. Therefore, to consider the whole features of the Wiener Brownian motion of the quantum states on $M$ one must incorporate the Ricci flow in the calculations to produce a dynamical background geometry. Actually, we should assume that the space metric would evolve with the Ricci flow. By definition the Ricci flow is:
\begin{equation} \label {b-2}
\frac{\partial}{\partial t} g_{ij}(t)=-2 Ric_{ij}(t),
\end{equation}

\noindent for $Ric(t)$ the Ricci curvature of evolving metric $g(t)$. In fact, as we will argue in the following, we regard the Ricci flow as an imposed natural entropic gauge fixing term for Einstein's field equation. One should note that although the Einstein equation is scale invariant, i.e. is symmetric with respect to transformation $g_{ij} \to c g_{ij}$ for some $c>0$, the Ricci flow disobeys this property. Hence, once the scale of time dimension $t$ is fixed\footnote{See Eq. (\ref {b-2}) in below where the time scale $\zeta$ is introduced.} the Ricci flow fixes the scale of $g_{ij}$ and consequently the space volume. Here we should point out that the Ricci flow has recently attracted an immense amount of attention in building a well-defined model of (topological) quantum gravity and several research papers have shown some theoretical evidence for an intimate correlation between gravity and the Ricci flow.\footnote{See for example \cite{top1, top2, top3, isi, graf, lulli, woolgar, dz, dz2, dz3, dowker, morison}.}

\par Actually, the Ricci flow has been established to be a very fruitful mechanism to improve metrics in Riemannian geometry whenever $M$ is compact. Indeed, it has been shown that in some specific conditions, the Ricci flow converges to a canonical metric \cite{hamilton1, hamilton2, hamilton3, huisken}. For instance, in his seminal work, Hamilton has established that on each closed 3-manifold with positive Ricci curvature the Ricci flow converges to a canonical metric of positive constant sectional curvature \cite{hamilton1}. In this sense, the Ricci flow may be regarded as a natural homotopy between a given metric of positive Ricci curvature and a canonical metric of constant sectional curvature. Moreover, Hamilton demonstrated that in this case the Ricci flow could be resolved within a normalized version of which the volume of the manifold is preserved along with the metric evolution. In principle, whenever the Ricci flow holds and admits a unique solution on $(a,b)\subset \mathbb R$, it converges to the canonical metric $g_c$ as $t \to b$, where $g_c$ is necessarily an Einstein metric, i.e., $\lim_{t\to b}Ric_{ij}(t)= K {g_{c}}_{ij}$ for some constant $K$.\footnote{For $K=\frac{1}{D}\Bigl( \int_M \lim_{t\to b} R(t) d\Omega_{g_c}/ \int_M d\Omega_{g_c} \Bigr)$, where $R$ is the Ricci scalar.}

\par Without loss of generality we may assume that the Ricci flow exists on $(-a,a)\subset \mathbb R$, for some $a>0$. Hence, by rescaling $t \mapsto t/n$, $n \in \mathbb N$, the domain of the solution changes to $(-na,na)$, whereas in its turn on the right-hand side of (\ref {b-2}) one must replace $-2$ with $-2/n$. In principle, since we are concerned about the stochastic process in $(-T,T)$ for $T\to \infty$, the integer $n$ must be chosen as large as we need. Generally, the Ricci flow could be rescaled as
\begin{equation} \label {b-2}
\frac{\partial}{\partial t} g_{ij}(t)=-2\zeta Ric_{ij}(t),
\end{equation}

\noindent for some small positive constant $\zeta$. In the following, we will see that as a very tiny amplitude, $\zeta$ couples to negligible terms inserted in the Einstein-Hilbert Lagrangian density. More precisely, it is seen that the Ricci flow scaling factor could be compared with the cosmological constant $\Lambda$.

\par In addition, it can be easily seen that the variation of the Riemannian volume form $d\Omega_{g(t)}$ with respect to $t$ is given in terms of the scalar curvature $R(t)$:
\begin{equation} \label {b-3}
\frac{\partial}{\partial t} d\Omega_{g(t)}=-\zeta R(t) d\Omega_{g(t)}.
\end{equation}

\noindent The above equation lets us work out the explicit formulation of the Wiener fractal measure (\ref {b-1}) along with the Ricci flow evolution of the space geometry. Initially, one should note that we have already considered a normalization condition as:
\begin{equation} \label {a-1}
\int_M \psi_{\lambda,i}(t)\psi_{\lambda',j}(t) d\Omega_{g(t)}=\delta_{\lambda \lambda'} \delta_{ij},
\end{equation}

\noindent where $\psi_{\lambda,i}(t)$ has been introduced in \textbf{Theorem 2}. Hence, by variation of (\ref {a-1}) with respect to $t$, $0<t \ll 1$, we obtain:
\begin{equation} \label {a-2}
\begin{gathered}
\epsilon_{\lambda, ii}(t)=1-\frac{t\zeta}{2}\int_M R(t) \psi_{\lambda,i}^2(t) d\Omega_{g(t)},\\
~~~~~~~~~~~~~~~~~\epsilon_{\lambda, ij}(t)+\epsilon_{\lambda, ji}(t)=-t\zeta \int_M R(t) \psi_{\lambda,i}(t)\psi_{\lambda,j}(t) d\Omega_{g(t)},~~~~~~~~~~(i\ne j)
\end{gathered}
\end{equation}

\noindent wherein
\begin{equation} \label {saya}
\epsilon_{\lambda, ij}(t)=\int_M \psi_{\lambda,i}(0)\psi_{\lambda,j}(t) d\Omega_{g(t)}.
\end{equation}

\noindent Hence, we would readily set;
\begin{equation} \label {a-4}
\epsilon_{\lambda, ij}(t)=\delta_{ij}-\frac{t\zeta}{2}\int_M R(t) \psi_{\lambda,i}(t)\psi_{\lambda,j}(t)d\Omega_{g(t)}.
\end{equation}

\noindent Consequently, along the variation of $g$ the basis of $\mathbf C_{\mathcal N}$ moves linearly by the matrix $\epsilon=\oplus_{\lambda} \epsilon_\lambda$, for all $\lambda \leq \mathcal N$, which means that the Fourier-Laplace coefficients $f_{\lambda,i}$s in (\ref {4}) must be transformed via $\epsilon^{-1}$. Therefore, in (\ref {b-1}) we must consider $F_i$ to be block diagonal for the spectrum of the Laplacian operator $\Delta^{g(t_{i-1})}$ as $F_i=\oplus_\lambda (F_i)_{\lambda}$, wherein
\begin{equation} \label {a-5}
~~~~~~~~~~~~~~~(F_i)_{\lambda,mn}=\delta_{mn} + \frac{\tau_i \zeta}{2}\int_M R(t_i)\psi_{\lambda,m}(\tau_i) \psi_{\lambda,n}(\tau_i) d\Omega_{g(t)},~~~~~~~~~~(1\leq i \leq N)
\end{equation}

\noindent for $m$ and $n$ running through the degenerate eigenfunctions of $\Delta^{g(t_{i-1})}$ with common eigenvalue $\lambda$ and $\{\psi_{\lambda,m}(0)\}$ is the basis of eigenfunctions of the Riemannian metric $g(t_{i-1})$. Indeed, one can simply consider (\ref {a-5}) as $F_i=1+r_i$, $i=1,\cdots, N$, where we easily gain;
\begin{equation} \label {a-6}
\begin{gathered}
J_i=\det F_i= \exp \left( tr \Big\{  \ln(1+r_i)  \Big\} \right)\approx \exp \left( tr \{r_i\} \right)\\
=\exp \left( \sum_{{\lambda \in \text{spec} \Delta^{g(t_{i-1})}}} \sum_{m\in D_\lambda}  \frac{\tau_i \zeta}{2} \int_M R(t_i) \psi^2_{\lambda,m}(\tau_i) d\Omega_{g(t_i)} \right),
\end{gathered}
\end{equation}

\noindent which is a local formula.

\par Incorporating the full expression of the Jacobian determinant (\ref {a-6}) in the Wiener fractal measure (\ref {b-1}) leads to a highly complicated formulation for Brownian motion of quantum states in $\mathbf C_{\mathcal N}$. Therefore, we have to consider our first approximation of the Wiener fractal measure as:\\

\par \textbf{Approximation I:} \emph{The double summation $\sum_{{\lambda \in \text{spec} \Delta^{g(t_{i-1})}}} \sum_{m\in D_\lambda}$ in the exponent of $J_i$ would be replaced by the dominating leading term $\sum_{\lambda=0}\sum_{m\in D_0}$. On the other hand, since $M$ is connected we have $\dim H^0_{dR}(M,\mathbb R)=1$. Hence, there is no degeneracy, i.e., $\dim D_0=1$, and the only normalized eigenfunction of $\Delta^{g(t_{i-1})}$ is a constant function. In fact, upon the normalization condition (\ref {a-1}) we obtain: $\psi_0=1/\sqrt{V_i}$, wherein $V_i$ is the volume of $M$ for the metric $g(t_i)$, i.e. $V_i=\int_M d\Omega_{g(t_i)}$. Therefore, we readily obtain:}
\begin{equation} \label {a-8}
J_i=\exp \left( \frac{\tau_i \zeta}{2V_i} \int_M R(t_i) d\Omega_{g(t_i)} \right).
\end{equation}
~
\par The second approximation concerns about the volume of the space manifold $M$. According to (\ref {b-3}) one can simply compute the derivation of this volume with respect to $t$ as:
\begin{equation} \label {a-8'}
\frac{dV(t)}{dt}=-\zeta\int_M R(t) d\Omega_{g(t)}.
\end{equation}

\noindent In fact, for any $t_i \in [-T,T]$, we compute:
\begin{equation} \label {a-8''}
V(t_i)=V -\zeta \int_{\mathcal M_{t_i}} R(t) dt\wedge d\Omega_{g(t)},
\end{equation}

\noindent wherein $V=V(-T)$ is the \emph{initial volume of the universe} and $\mathcal M_{t_i}=[t_i,-T]\times M$ is the \emph{truncated space-time continuum}. Since $\zeta \approx 0$ we could assume that $V(t)$ evolves slightly, provided the integral of the scalar curvature of the space manifold $M$ is not too huge. As we see from (\ref {a-8''}) the coefficient $1/V_i$ is itself given by an integration over the truncated space-time continuum $\mathcal M_{t_i}$. Therefore, inserting (\ref {a-8''}) in (\ref {a-8}) results in a highly non-local formula for the gravity effects. Thus, the next approximation is reasonable:\\

\par \textbf{Approximation II:} \emph{To avoid the non-local contributions we must ignor the variation of $V_i$ along with the Ricci flow. Consequently, we would simply set $V_i=V$, $1\leq i \leq N$, and write the Jacobian determinant (\ref {a-8}) as:}\footnote{Indeed, as we see in the following the space volume $V$ is effectively equivalent to the Newton's gravitation constant $G$. In principle, upon (\ref {a-8''}) the dynamics of the space volume is formulated by a double integral on $M\times M$ within the Einstein-Hilbert action, hence produces a non-local contribution of the scalar curvature of $M$ into the theory. Therefore, the small variation in the volume of the universe $V$, and consequently in the gravitational constant $G$, is the only mandatory choice we have to consider in the first order approximations of the Wiener fractal measure. However, the slight variation of $G$ has already been formulated by means of a local interacting scalar field theory (i.e. dilaton) via the Brans-Dicke theory \cite{brans-dicke}. See also \cite{weinberg} for more discussions.}
\begin{equation} \label {a-9}
J_i=\exp \left( \frac{\tau_i \zeta}{2V} \int_M R(t_i) d\Omega_{g(t_i)} \right).
\end{equation}
~
\par In particular, upon approximations \textbf{I} and \textbf{II} one can easily extract the contribution of the Jacobian determinants $J_i$s, $1\leq i \leq N$, within the Wiener fractal measure (\ref {b-1}) as
\begin{equation} \label {a-10}
\prod_{i=1}^N J_i =\exp \left(  \frac{1}{2 \xi} \sum \tau_i \int_M R(t_i) d\Omega_{g(t_i)} \right)
\end{equation}

\noindent for $\xi = V/\zeta$, which in the limit of $N \to \infty$ (i.e. $\tau_i =2T/N \to 0$) leads to:
\begin{equation} \label {a-10}
\prod_{i=1}^N J_i= \exp \left( \frac{1}{2  \xi c} \int_{\mathcal M} R~d\Omega \right),
\end{equation}
 
\noindent wherein $\mathcal M=[-T,T]\times M$ is the space-time continuum introduced above, and $d\Omega=c dt\wedge d\Omega_{g(t)}$ is the Riemannian volume form for the Lorentzian metric $\mathbf g=\mathbf g_{\mu \nu} dx^\mu \otimes dx^\nu= dx^0 \otimes dx^0 \oplus \left(-g_{ij}(t)dx^i \otimes dx^j\right)$ on $\mathcal M$. We should note that according to the rescaled Ricci flow (\ref {b-2}) the scalar curvature $R$ of $g$ is essentially equal to the scalar curvature $\mathcal R$ of $\mathbf g$ with an overall minus sign and some additional negligible terms of order $\mathcal O(\zeta)$. In fact, upon to (\ref {b-2}) we see:
\begin{equation} \label {a-12}
\mathcal R=-R+\frac{\zeta}{c^2} \Big\{g^{ij} \frac{\partial Ric_{ij}}{\partial t} + \frac{\partial R}{\partial t}\Big\} -\frac{\zeta^2}{c^2} Ric_{ij}Ric^{ij}+\frac{\zeta^2}{c^2} R^2=-R+\mathcal O(\zeta).
\end{equation}

\noindent Here, we may set our third approximation law:\\

\par \textbf{Approximation III:} \emph{We would simply replace $R$ by $-\mathcal R$ in the Jacobian determinant $J_i$ and ignore the remaining terms of order $\mathcal O(\zeta)$. Thus, we obtain:}
\begin{equation} \label {a-13}
\prod_{i=1}^N J_i= \exp \left( - \frac{1}{2  \xi c} \int_{\mathcal M} \mathcal R~d\Omega \right).
\end{equation}
~
\par In principle, the above three approximations would provide a sufficient set of assumptions to work out the original formulation of general relativity via the Einstein-Hilbert action. Actually, in a local coordinate system on the space-time continuum $\mathcal M$, say $(t,x^i)$, we would obtain the following formula for the Jacobian term (\ref {a-13}):
\begin{equation} \label {a-11}
\prod_{i=1}^N J_i=\exp \left( - \frac{1}{2 \xi c} \int_{\mathcal M}\sqrt {|\det \mathbf g|} ~\mathcal R~ dx^0\wedge dx^1 \wedge \cdots \wedge dx^D \right).
\end{equation}

\noindent Hence, the Wiener stochastic process of quantum states on the Riemannian manifold $M$ along with the Hamilton-Perelman Ricci flow automatically involves the Einstein-Hilbert action of general relativity via the approximations \textbf{I}, \textbf{II}, and \textbf{III}. In fact, in the above formulations, the emergence of general relativity is merely due to the evolution of the cosmos geometry, hence having nothing to do with possible background interactions of the included matter fields. However, one should note that the analytic structure of the spatial metric $g_{ij}$ fulfills Einstein's field equation via the Euler-Lagrange equations. In other words, although $\mathbf g_{\mu \nu}$ is affected by the energy-momentum tensor of the matter fields via the optimization of the total action, it must obey the Ricci flow as the imposed gauge fixing term that determines the intrinsic evolution of the space-time geometry.

\par Following the machinery of the previous section, it can be easily seen that the rest parts of the Wiener fractal measure on the dynamical Riemannian manifold $(M,g(t))$ would be
\begin{equation} \label {aa-1}
 \exp \left( - \frac{1}{\hbar} \int_{\mathcal M} \frac{1}{2}\Big\{ \mathbf g^{\mu \nu }\partial_\mu \phi \partial_\nu \phi - \frac{m^2c^2}{\hbar^2} \phi^2 \Big\} d\Omega \right) \frak D \phi,
\end{equation}

\noindent wherein the strategy \textbf{A} is considered accordingly. One should note that there is an essential difference between Wiener measures (\ref {13}) and (\ref {aa-1}). In the former, the space-time metric $\mathbf g$ is a static tensor and produces no gravitational effect, but for the latter, it evolves via the Ricci flow and admits Einstein's field equation. Moreover, in the latter, the Feynman measure $\frak D \phi$ is given in terms of the Fourier expansion of the Laplacian operator $\Delta^{g(t_{i-1})}$ at the $i$-th time section, whereas in the former the Feynman measure is inherently static. All in all, the Wiener fractal measure along with the Ricci flow and after implementing the mentioned three approximations and the strategy \textbf{A} becomes
\begin{equation} \label {aa-2}
\begin{gathered}
dW(t_N,\cdots,t_1)\approx  \exp \left(  -\frac{1}{\hbar} \int_{\mathcal M} \Big\{\frac{1}{2 \kappa} \sqrt{|\det \mathbf g|}~\mathcal R  + \mathcal L_{K-G} (\phi) \Big\} dx^{D+1} \right) \frak D \phi,
\end{gathered}
\end{equation}

\noindent with $\mathcal L_{K-G}(\phi)=\frac{\sqrt{|\det \mathbf g|}}{2}\big\{ \mathbf g^{\mu \nu }\partial_\mu \phi \partial_\nu \phi - \frac{m^2 c^2}{\hbar^2} \phi^2 \big\} $ the Klein-Gordon Lagrangian density, and $\kappa =\xi c/\hbar$. Hence, we readily obtain Newton's gravitational constant in terms of fractal, geometric and physical parameters in $D=3$ dimensions: $G=\xi c^4/8\pi \hbar$. Therefore, if we use the standard values of $G$, $\hbar$ and $c$ we compute: $\xi=V/\zeta=2.1899700 \times 10^{-77}~(m.s)$. This shows that the initial volume of the universe $V$ (i.e. the cosmos volume at the initial point at which the gravitational effects have emerged) has been extremely small. Indeed, the smallness of the cosmos' initial volume $V$ is, in fact, equivalent to a large amount of the Einstein-Hilbert coefficient (i.e. $1/2\kappa=c^4/16\pi G$) in the SI units. Thus, the extremely small amount of the cosmos' initial volume $V$, as we will explain later, could be regarded as the main reason for the hierarchy problem.\footnote{See \cite{hier} and the references therein for interesting discussions about the relation of the hierarchy problem and the Bayesian statistics. It should be noted that in this article we also have practically found a relationship between the hierarchy problem and the Bayesian statistics.}

\par To work out a well-defined theory from the Wiener fractal measure (\ref {aa-2}) we must employ the strategy \textbf{B} once again. We readily obtain:
\begin{equation} \label {aa-3}
\begin{gathered}
dW(t_N,\cdots,t_1)\approx  \exp \left(  \frac{i}{\hbar} \int_{\mathcal M} \Big\{\frac{1}{2 \kappa} \sqrt{|\det \mathbf g|}~\mathcal R  + \mathcal L_{K-G} (\phi) + \sqrt{|\det \mathbf g|}~ \frac{i\varepsilon}{2} \phi^2 \Big\} dx^{D+1} \right) \frak D \phi,
\end{gathered}
\end{equation}

\noindent for $\varepsilon \to 0$. Therefore, one can consider the Wiener fractal measure (\ref {aa-3}) obtained by implementing the Wiener stochastic process (\ref {b-1}) of the Lebesgue fractal measure (\ref {4}) through with the Ricci flow as the best local approximation of the more fundamental formulation of quantum field theory in dynamical curved spacetime, i.e., in the presence of semi-classical gravity. Hence, based on the above results it seems that the basic source of Einstein's theory of gravity stems from the dynamical Wiener fractal measure via the Ricci flow. We refer to this stochastic derivation of gravitational effects as the \emph{Wiener fractal gravity}. As we discussed in the last section, although including non-local terms the Wiener fractal gravity leads to finite predictable expectation values such as
\begin{equation} \label {aa-4}
\lim_{\mathcal N, N \to \infty} \int_{\mathbf C_\mathcal N} F^{(\mathcal N)}(t_N,\cdots,t_1) \big\{\phi(t_{i_1}) \cdots \phi(t_{i_n})   \big\}dW(t_N,\cdots,t_1)
\end{equation}

\noindent for $F^{(\mathcal N)}(t_N,\cdots,t_1)=\prod_{k=1}^N \exp \left( \tau_k \int_M V^{(\mathcal N)} (\phi(t_{k-1})) d\Omega_{g(t_{k-1})}\right)$ with negative (or bounded from above) and renormlizable interaction terms $V^{(\mathcal N)}(\phi(t_i))$. This ensures the significance of the Wiener fractal gravity as the fundamental and genuine formulation of Feynman's path-integral measure for quantum field theory in the presence of semi-classical gravitational effects. In principle, considering the Wiener process of propagating quantum states on $M$ and the involved entropic/geometric features of the second law of thermodynamics via the Hamilton-Perelman Ricci flow will automatically lead us to the Wiener fractal gravity. Through the next two sections, we try to study the theoretical and empirical properties of the Wiener fractal gravity.

\par
\section{Ricci Flow, Cosmological Constant and $\Lambda$CDM Model}
\setcounter{equation}{0}

\par Upon Hamilton's celebrated theorem we may consider the metric of the universe to be almost Einstein with $Ric_{ij}(t)=K g_{ij}(t)+\varepsilon_{ij}$, for some infinitesimal symmetric matrix $\varepsilon_{ij}\ll Kg_{ij}$ and the cosmological time $t\geq t_0$, where $t_0$ could be considered as the current age of the universe.\footnote{In fact, according to experimental achievements, $K$ has a very tiny amplitude; i.e. $K \approx 0$. Indeed, the experimental data from different laboratories (for example WMAP, BOOMERanG, and Planck) confirm that the universe manifold $M$ is flat with about $0.4$ percent margin of deviation \cite{yoo}. See also \cite{dem, levin}.} With this assumption, the Ricci flow is;
\begin{equation} \label {a-14}
~~~~~~~~~~~~~~~\frac{\partial g_{ij}(t)}{\partial t}=-2\zeta K g_{ij}(t)-2\zeta \varepsilon_{ij},~~~~~~~~~~(t \geq t_0)
\end{equation}

\noindent which has a unique solution as\footnote{We may temporarily assume that $\varepsilon_{ij}$ is constant.}
\begin{equation} \label {a-15}
g_{ij}(t)=e^{-2\zeta K (t-t_0)} \Bigl( g_{ij}(t_0) + \frac{\varepsilon_{ij}}{K} \Bigr) - \frac{\varepsilon_{ij}}{K}, ~~~~~~~~~~Ric_{ij}(t)=K e^{-2\zeta k (t-t_0)} \Bigl( g_{ij}(t_0) + \frac{\varepsilon_{ij} }{K} \Bigr).
\end{equation}

\noindent Hence, the higher order terms of (\ref {a-12}) would become;
\begin{equation} \label {a-16}
\begin{gathered}
\frac{\zeta }{c^2} \Big\{g^{ij} \frac{\partial Ric_{ij}}{\partial t} + \frac{\partial R}{\partial t}\Big\}- \frac{\zeta^2}{c^2} Ric_{ij}Ric^{ij}+ \frac{ \zeta^2}{c^2} R^2\\
= (D-3) D c^{-2} \zeta^2 K^2+ 2(D-3)\zeta^2 c^{-2} K e^{-2\zeta k (t-t_0)} \Bigl( tr\{ \varepsilon g_0 \} + tr \{ \varepsilon^2 / K \} \Bigr) +2 \Lambda,
\end{gathered}
\end{equation}

\noindent for $g_0=g(t_0)$, and the cosmological constant $\Lambda$:\footnote{Actually, $\Lambda$ is not a constant in general, but upon to Hamilton's theorem \cite{hamilton1} for some appropriate interval of time it evolve slightly on $M$, hence could be considered as a constant at the first order of approximation.}
\begin{equation}
\Lambda=\frac{1}{2}\zeta^2 c^{-2} e^{-4\zeta K (t-t_0)} \Big\{ tr \{\varepsilon g_0 \}^2 -tr \{ \left( \varepsilon g_0 \right)^2 \}\Big\} + \mathcal O (\varepsilon^3 ),
\end{equation}

\noindent with
\begin{equation} \label{april-0}
 \mathcal O (\varepsilon^3 ) = \frac{1}{2}\zeta^2 c^{-2} e^{-4\zeta K (t-t_0)} \Bigg\{ 2 \Bigl( tr \{\varepsilon g_0 \} tr \{\varepsilon^2/K \} - tr \{ g_0 \varepsilon^3/K  \} \Bigr) + \Bigl(  tr \{\varepsilon^2/K \} ^2 - tr \{ \bigl( \varepsilon^2/K \bigr)^2 \} \Bigr) \Bigg\},
\end{equation}
 
\noindent which could be neglected at first glance in our calculations. 

\par Actually, upon Hamilton's theorem \cite{hamilton1}, the deviation matrix $\varepsilon$ is positive definite for $D=3$ dimensions, hence we conclude that the cosmological constant $\Lambda$ is positive on closed 3-manifolds. To see this, we may ignore the contribution of (\ref {april-0}) in $\Lambda$ and rewrite it as
\begin{equation} \label{april-1}
\Lambda=\frac{1}{2}\zeta^2 c^{-2} e^{-4\zeta K (t-t_0)} \Big\{ tr \{\overline \varepsilon  \}^2 -tr \{ \left( \overline \varepsilon \right)^2 \}\Big\},
\end{equation} 

\noindent for $\overline \varepsilon = g_0^{1/2} \varepsilon g_0^{1/2}$. Here we must note that the matrix product and $tr$ in (\ref {april-1}) are respectively the algebraic matrix product and the algebraic trace without the intervention of the metric or the contraction of the lower and upper indices, i.e. we have:
\begin{equation} \label{april-2}
(\varepsilon g_0)_{ij}=\sum_{k=1}^D \varepsilon_{ik}{g_0}_{kj},~~~~~tr\{\varepsilon g_0\} = \sum_{i=1}^D(\varepsilon g_0)_{ii},~~~~~(\varepsilon g_0)^2_{ij}=\sum_{k=1}^D(\varepsilon g_0)_{ik}(\varepsilon g_0)_{kj},~~~~~tr\{\left( \varepsilon g_0 \right)^2\}=\sum_{i=1}^D \left( \varepsilon g_0 \right)^2_{ii}.
\end{equation}

\noindent Hence, to prove the positivity of $\Lambda$ we may avoid employing coordinate transformations. Nevertheless, since $\varepsilon$ is positive definite in $D=3$ dimensions and $\varepsilon$ and $g_0^{1/2}$ are both symmetric, then $\overline \varepsilon$ is itself a diagonalizable positive definite matrix with three positive eigenvalues $\lambda_1$, $\lambda_2$, and $\lambda_3$. Thus, we obtain:
\begin{equation} \label {april-3}
\Lambda=\zeta^2 c^{-2} e^{-4\zeta K (t-t_0)} \left( \lambda_1 \lambda_2 + \lambda_1 \lambda_3 + \lambda_2 \lambda_3 \right), 
\end{equation}

\noindent which is obviously positive.

\par On the other hand, we see surprisingly that in $D=3$ dimensions the two first leading terms of (\ref {a-16}) vanish and we only obtain the constant term $2\Lambda$ on the right-hand side of the equation. Thus, on the three-dimensional space $M$ we readily find $R=-\mathcal R+2 \Lambda$. Consequently, the Jacobian term (\ref{a-11}) must be revised via this precision, and the cosmological constant term $-2\Lambda$ has to be simply inserted beside the scalar curvature $\mathcal R$ within the Einstein-Hilbert Lagrangian density. Strictly speaking, we will obtain:
\begin{equation} \label {a-17}
\prod_{i=1}^N J_i=\exp \left( - \frac{1}{2\xi c} \int_{\mathcal M}\sqrt {|\det \mathbf g|} ~\left(\mathcal R-2\Lambda \right)~ dt\wedge dx^1 \wedge \cdots \wedge dx^D \right).
\end{equation}

\noindent Therefore, according to the above calculations, the cosmological constant emerges only upon the metric evolution on the space manifold. In other words, the source of the cosmological constant in the Wiener fractal gravity is purely geometric/entropic and independent from the matter fields or the zero-point energy fluctuations of the quantum field theory. That is, the cosmological constant $\Lambda$ has an entropic source due to the Ricci flow of the three-dimensional geometry of the universe and is not an inevitable consequence of the matter fields and their interactions. In addition, since $\Lambda \propto e^{-4\zeta K t} \zeta^2 \mathcal O(\varepsilon^2)$, the cosmological constant has essentially an extremely small amplitude.\footnote{In principle, here $\Lambda$ is itself a field on space-time manifold. Nevertheless, to align the Wiener fractal gravity with the standard formulation in the literature, we should make an approximation and insert the mean value of $\Lambda$ in the Enstein-Hilbert action in (\ref {a-17}).} Consequently, it seems that the Wiener fractal gravity would provide a new insight to the cosmological constant problem \cite{hobson, weinberg-cos} and the nature of dark energy.\footnote{See \cite{cai} for another entropic interpretation of the cosmological constant via the holographic approach in quantum gravity.} 

\par Moreover, for almost canonical asymptotic Einstein metric $g_c$ of the Hamilton-Perelman Ricci flow, i.e. $Ric_{ij}\approx K {g_c}_{ij}$, we readily obtain an inflating universe $M$. To see this one has to employ the Jacobi field equation for Einstein metric $g_c$. Let $\gamma_{s}(t)$ be a continuous family of geodesics on $\mathcal M$ labeled by $s \in \mathbb R$, each of which with initial condition $\dot{\gamma}_s(0)=\partial/\partial t$. Then, according to the Jacobi field equation we read for small enough $t$:
\begin{equation} \label {a-18}
\frac{D^2}{dt^2}J(t)=\mathcal R\left(\dot{\gamma}_s(t),J(t)\right)\dot{\gamma}_s(t) \approx {\mathcal R_{0i0}}^jJ^i(t) \frac{\partial}{\partial x^j},
\end{equation} 

\noindent wherein $J(t)=\frac{\partial \gamma_s(t)}{\partial s}$ and $D/dt$ is the covariant derivative along $t$, while ${\mathcal R_{\mu \nu \sigma}}^\lambda$ is the space-time Riemann curvature tensor computed for the metric $\mathbf g=c^2 dt\otimes dt \oplus (-g_c)$.\footnote{Here the coordinate index $\mu$ varies among $\{0,1,\cdots,D\}$, where $\mu=0$ corresponds to the cosmological time $t$.} Let us compute the relative acceleration of inertial celestial bodies due to the space-time metric $\mathbf g$. We simply set $J(0)=r\frac{\partial}{\partial x^i}$, for some fixed $i \in \{1,\cdots, D\}$, assuming the observed galaxies are located along the $i$-th direction of the local coordinate system $(x^\mu)=(x^0=t,x^1,\cdots,x^D)$ with $\gamma_s(0)=(0, \cdots, 0, sr, 0, \cdots,0)$ for some $r=r(0)>0$ and $s \in [0,1]$. Moreover, upon the Ricci flow (\ref {b-2}) we have:
\begin{equation} \label {chris - ricci}
\Gamma^0_{ij}=\zeta Ric_{ij},~~~~~\Gamma^i_{0j}=\Gamma^i_{j0}=- \zeta {g}^{im}Ric_{mj},~~~~~\Gamma^0_{0i}=\Gamma^0_{i0}=\Gamma^i_{00}=\Gamma^0_{00}=0.
\end{equation}

\noindent However, by assuming the asymptotic Einstein metric $g_c$ we have: $Ric_{ij}\approx K {g_c}_{ij}$. Thus, we obtain:
\begin{equation} \label {chris - ricci - eins}
\Gamma^0_{ij}=\zeta K g_{cij},~~~~~\Gamma^i_{0j}=\Gamma^i_{j0}=-\zeta K \delta^i_j,~~~~~\Gamma^0_{0i}=\Gamma^0_{i0}=\Gamma^i_{00}=\Gamma^0_{00}=0.
\end{equation}

\noindent Therefore, we compute the relative acceleration as 
\begin{equation} \label {a-19}
\begin{gathered}
a_r=a_r^\mu \frac{\partial}{\partial x^\mu}={\mathcal R_{0i0}}^\mu \frac{\partial}{\partial x^\mu}=r\Bigg\{\frac{\partial}{\partial t}\Gamma^\mu_{0i}-\frac{\partial}{\partial x^i}\Gamma^\mu_{00}+\Gamma^\mu_{0\nu}\Gamma^\nu_{i0}-\Gamma^\mu_{i \nu}\Gamma^\nu_{00} \Bigg\} \frac{\partial}{\partial x^\mu}=\zeta^2 K^2 r \frac{\partial}{\partial x^i},
\end{gathered}
\end{equation}

\noindent which is, in fact, equal to:
\begin{equation} \label {a-20}
~~~~~~~~~~~~~~~\ddot{r}(t)=H_\infty^2 r(t),~~~~~~~~~~(t\geq 0)
\end{equation}

\noindent for radial distance $r(t)$ and \emph{ultimate Hubble's constant}:\footnote{According to Hamilton's theorem \cite{hamilton1} $K$ is considered to be a positive number. However, here we may also consider negative $K$ to involve Einstein manifolds with negative curvature. We need this generalization in the FLRW metric. Hence, in (\ref {H_infty}) we use the absolute value of $K$.}
\begin{equation} \label {H_infty}
H_\infty=\zeta |K|.
\end{equation}

\noindent In principle, (\ref {a-20}) has a unique dominating solution in terms of the cosmological time $t$ as
\begin{equation} \label {a-21}
~~~~~~~~~~~~~~~~~~~~r(t)=r_0 e^{H_\infty(t-t')},~~~~~~~~~~(t\geq t')
\end{equation}

\noindent wherein we assume the initial time $t'$ to be close enough to the ultimate time for the Ricci flow to reach the promised Einstein's metric $g_c$.

\par These conclusions could be slightly modified via the assumption of (\ref {a-14}) for the present age of the universe $t_0$ and the upcoming future of the cosmological time $t>t_0$. In fact, for this general case, we consider the two following equations:
\begin{equation} \label {jadid-1}
\frac{\partial}{\partial t} g_{ij}(t) = -2 \zeta Ric_{ij}(t),~~~~~\text{and} ~~~~~ Ric_{ij}(t)=K g_{ij}(t) + \varepsilon_{ij}(t),
\end{equation}

\noindent for $t\geq t_0$ and $\varepsilon_{ij}(t) /K \ll g_{ij}(t)$. We can obtain the metric $g_{ij}(t)$ and its inverse $g^{ij}(t)$ accordingly:
\begin{equation} \label {jadid-2}
\begin{gathered}
g_{ij}(t) = e^{-2K\zeta (t-t_0)} \Bigl( g_{ij}(t_0) + \epsilon_{ij} (t) \Bigr) ~~~~~~~~~~~~~~~~~
 g^{ij}(t) =  e^{2K\zeta (t-t_0)} \Bigl( g^{ij}(t_0) - \epsilon^{ij} (t) \Bigr),
 \end{gathered}
\end{equation}

\noindent wherein
\begin{equation} \label {jadid-3}
\epsilon_{ij}(t) = -2 \zeta \int_{t_0}^t e^{2 K\zeta (t' - t_0)} \varepsilon_{ij}( t' ) dt'
\end{equation}

\noindent and $\epsilon^{ij} (t) = g^{im}(t_0) \epsilon_{mn} (t)  g^{nj}(t_0)$. Hence, we readily find:
\begin{equation} \label {jadid-4}
\Gamma^i_{0j}=\Gamma^i_{j0} = - \zeta K \delta^i_j  - \zeta e^{2 K \zeta (t - t_0)} \Bigl( {A^j_0}_i(t) - \epsilon^{im}(t) \varepsilon_{mj}(t) \Bigr).
\end{equation}

\noindent for ${A^j_0}_i(t) =g^{jm}(t_0) \varepsilon_{mi}(t) $. Therefore, in this case, we obtain the accurate form of (\ref {a-19}) as\footnote{As we see once again the $\varepsilon$-linear term cancels out automatically.}
\begin{equation} \label {a-22}
\begin{gathered}
a_r=a_r^\mu \frac{\partial}{\partial x^\mu}=\\
\zeta^2 K^2 r \frac{\partial}{\partial x^i} - \zeta^2 r e^{4 K \zeta (t - t_0)} A^i_{0k}(t)  A^k_{0j} (t)  \frac{\partial}{\partial x^j} - \zeta e^{2K\zeta (t-t_0)}  \mathcal A^j_i(t) r \frac{\partial}{\partial x^j}+\mathcal O(\varepsilon^3),
\end{gathered}
\end{equation}

\noindent with
\begin{equation} \label {a-23}
\mathcal A^j_i (t)= \Bigl(  \epsilon^{im} (t) - g^{im}(t_0) \Bigr) \dot{\varepsilon}_{mj}(t).
\end{equation}

\noindent The dominating diagonal term of (\ref {a-23}) (for $\varepsilon_{ij}(t) \approx \varepsilon (t) \delta_{ij}$ and $g_0^{ij} \approx \delta^{ij}$) is:
\begin{equation} \label {a-24}
\begin{gathered}
a_r=a_r^\mu \frac{\partial}{\partial x^\mu}= r \Bigg\{ \zeta^2 K^2 - \zeta^2 \varepsilon^2(t) e^{4K\zeta(t-t_0)} - \zeta \dot \varepsilon (t)  e^{2K\zeta (t-t_0)} ( \epsilon (t) -1 ) \Bigg\} \frac{\partial}{\partial x^i},
\end{gathered}
\end{equation}

\noindent for
\begin{equation} \label {jadid-5}
\epsilon(t) =  -2 \zeta \int_{t_0}^t e^{2K\zeta (t'-t_0)} \varepsilon (t') dt'.
\end{equation}

\noindent The above calculation leads to a time-dependent Hubble's parameter $H(t)$ as
\begin{equation} \label {a-25}
H(t)=H_\infty \sqrt{1 - \frac{\varepsilon^2 (t)}{K^2} e^{4K \zeta (t-t_0)} - \frac{ \dot \varepsilon (t)}{\zeta K^2} e^{2K\zeta (t-t_0)} \Bigl( \epsilon(t) -1   \Bigr) }.
\end{equation}

\noindent Upon Hamilton's theorem \cite{hamilton1} the integral of (\ref {jadid-3}) is convergent, hence $e^{2K \zeta (t - t_0)} \varepsilon (t)$ must be a decreasing function. Therefore, the time-dependent function $- \frac{\varepsilon^2 (t)}{K^2} e^{4K \zeta (t-t_0)}$ is increasing and $\dot \varepsilon (t) < -2K\zeta \varepsilon(t) <0$. Thus, according to (\ref {jadid-5}) the time-dependent function $- \frac{ \dot \varepsilon (t)}{\zeta K^2} e^{2K\zeta (t-t_0)} ( \epsilon(t) -1 )$ is increasing too. Set
\begin{equation} \label {a-25'}
\Omega_1(t)= -\frac{\varepsilon^2(t) }{K^2}  e^{4K \zeta(t-t_0)} ~~~~~~~~\text{and}~~~~~~~~\Omega_2(t)= - \frac{ \dot \varepsilon (t)}{\zeta K^2} e^{2K\zeta (t-t_0)} \Bigl( \epsilon (t) - 1  \Bigr).
\end{equation}

\noindent Therefore, we gain approximately\footnote{We are working in a period of time $[t_0,T]$ in which $\Omega_i(t_0) \Delta \Omega_j(t) \ll 1$, with $t_0 \leq t \leq T$, for $i,j=1,2$.}
\begin{equation} \label {a-26}
H(t)=H_0 \sqrt{ 1+  \Delta \Omega_1(t) + \Delta \Omega_2(t)  },
\end{equation}

\noindent for positive functions $\Delta \Omega_i(t)=\Omega_i(t)-\Omega_i(t_0)$, $i=1,2$, and the Hubble's constant $H_0$ as
\begin{equation} \label {a-27}
H_0=H_\infty \sqrt{ 1+\Omega_1(t_0) + \Omega_2(t_0)  }.
\end{equation}

\noindent All in all, we consequently find out the following major results: \\

\par \textbf{a)} \emph{The universe is expanding radially, hence radial velocities must be observed among extra-galactic nebulae and the amplitudes of these velocities must be proportional to distances from the fiducial observer, just as Hubble pointed out in his seminal 1929 paper \cite{hubble}.}

\par \textbf{b)} \emph{The acceleration is positive, hence galaxies flee from each other and the universe is inflating with an increasing rate of}\footnote{One should note that the inflation of the universe does not geometrically mean that the volume $V(t)$ of $M$ is increasing along the Ricci flow. Indeed, according to (\ref {b-3}) the volume of $M$ for Einstein metric $g_c$ changes with the rate of $dV(t)/dt=-\zeta K DV(t)<0$. Hence, for positive $K$ the volume is decreasing along with the Ricci flow, while practically the galaxies flee from each other.}
\begin{equation} \label {a-26'}
\frac{\dot a}{a}=H_0 \sqrt{ 1+  \Delta \Omega_1(t) + \Delta \Omega_2(t)  },
\end{equation}

\noindent \emph{where $a(t)=\left(1+z \right)^{-1}$ is the scale factor for observed redshift $z$.}\footnote{See \cite{mccann} as an intersting research which shows that the root-mean-square distance of two density volume terms $\omega_1(t)\geq 0$ and $\omega_2(t)\geq 0$ evolving with the heat (diffusion) equation $\frac{\partial}{\partial t}\omega_i=\Delta^{g(t)}\omega_i$, $i=1,2$, is non-increasing along the backward Ricci flow $\frac{\partial}{\partial t} g_{ij}=2 Ric_{ij}$. Indeed, the inverse of this claim via the transformations $t\to -t$ would lead to an expansion theorem along the Ricci flow for density fields. We regard this conclusion as an entropic force, which obviously includes the ordinary pressure in thermodynamics. This result, in principle, demonstrates the entropic root of the Ricci flow in its own turn.}

\par \textbf{c)} \emph{According to the arguments of the next section for similarities between the second law of thermodynamics and the Wiener fractal gravity, hereby we conclude that the dark energy (and the negative pressure due) stems from an entropic/geometric source due to the Ricci flow.}\footnote{See \cite{cai, chang} for more discussions about the entropic source of inflation.}\\

\par Indeed, the fundamental geometric assumptions of the Wiener fractal gravity resemble the well-understood Friedmann-Lamitre-Robertson-Walker solution, where Einstein's field equation is actually used to work out the scale factor $a(t)$ of the spatial metric
\begin{equation} \label {jadid-F}
g_{ij} dx^i \otimes dx^j=a^2(t)\left( \frac{dr \otimes dr }{1-kr^2}+r^2 \left(d\theta \otimes d\theta +\sin^2 \theta ~d\phi \otimes d\phi \right) \right)
\end{equation}

\noindent as a function of cosmological time $t$. This provides more common results between the $\Lambda$CDM model and the background geometric assumptions of the Wiener fractal gravity that includes the separation of space and time in the space-time continuum and the assumption of the Hamilton-Perelman Ricci flow for the spatial metric. In fact, for the spatial part of the FLRW solution in reduced-circumference polar coordinates (\ref {jadid-F}) we readily find: $Ric_{ij}=-2kg_{ij}$. Hence, $g_{ij}$ in (\ref {jadid-F}) is in its own turn an Einstein metric that is the promised asymptotic solution $g_{c ij}$ of the Ricci flow (for $K=-2k$):
\begin{equation} \label {a-227}
Ric_{ij}=K g_{c ij}.
\end{equation}

\noindent Thus, upon the above arguments, the FLRW metric would describe the geometric aspects of the universe in an admissible agreement with the background geometry of the Wiener fractal gravity.

\par However, if we readily impose the Ricci flow (\ref {b-2}) on the FLRW metric (\ref {jadid-F}), then we simply obtain a first order differential equation for the scale factor $a(t)$ as
\begin{equation} \label {jadid-FE}
a(t)\dot a(t) = 2 \zeta k a^2(t) ~~~~~\to ~~~~~\dot a(t)=2\zeta k a(t)=H_\infty a(t), ~~~~~(k=1)
\end{equation}

\noindent with the immediate solution
\begin{equation} \label {jadid-F0}
a(t)=a_0 e^{H_\infty (t-t_0)}.
\end{equation}

\noindent Nevertheless, based on Hamilton's theorem for 3-manifolds \cite{hamilton1}, the Einstein metric is the ultimate solution of the Ricci flow, hence the FLRW metric is indeed an asymptotic solution for the Wiener fractal gravity. Anyway, when the Ricci flow is employed this asymptotic solution leads to simple equalities of
\begin{equation} \label {jadid-F1}
\frac{\dot a}{a}=H_\infty ~~~~~ \text{and} ~~~~~ \frac{\ddot a}{a}=H_\infty^2.
\end{equation}

\noindent Based on (\ref {jadid-F1}) the Friedmann equations turn to:
\begin{equation} \label {a-27'}
H_\infty^2+\frac{kc^2}{a^2}-\frac{\Lambda c^2}{3}=\frac{8\pi G}{3} \rho~~~~~~~~~~\text{and}~~~~~~~~~~H_\infty^2+\frac{kc^2}{3 a^2}-\frac{\Lambda c^2}{3}=-\frac{8\pi G}{3c^2} p,
\end{equation}

\noindent which result in immediate asymptotic solutions;
\begin{equation} \label {a-30}
\begin{gathered}
\rho(t)=\frac{1}{8\pi G} \Big\{ 3 H_\infty^2 +c^2 \left( \frac{3 k}{a_0^2} e^{-2H_\infty (t-t_0)} -\Lambda \right) \Big\},\\
p(t)=\frac{c^2}{8\pi G} \Big\{ c^2 \left( \Lambda - \frac{k}{a_0^2} e^{-2H_\infty (t-t_0)} \right) -3 H_\infty^2 \Big\}.
\end{gathered}
\end{equation}

\par Before closing this section we prefer to return to (\ref {a-26}) to have a brief look at the Hubble's parameter $H(t)$. It is well-known that the Friedmann equation is rewritten in terms of the density parameters as
\begin{equation} \label {a-31}
H(t)=H_0\sqrt{ \Omega_{mat} a^{-3} + \Omega_{rad} a^{-4} + \Omega_k a^{-2} + \Omega_{\Lambda} a^{-3(1+\omega)} },
\end{equation}

\noindent wherein $\Omega_{mat}$ is the matter density including baryons and the cold dark matter, $\Omega_{rad}$ is for radiation including neutrinos,\footnote{Actually, the radiation has a negligible amplitude \cite{fri}. See also \cite{peebles} as a nice presentation of the issue.} $\Omega_k$ is due to the curvature, and $\Omega_{\Lambda}$ is the contribution of dark energy with the equation of the state parameter $\omega$. By imposing the Ricci flow on the Friedmann equations for positive curvature $k$ (i.e. $K<0$ in (\ref{a-227})), one simply reads from (\ref {a-25'}); $\Delta \Omega_1 \propto \varepsilon^2 a^{-4}$ and $\Delta \Omega_2 \propto \dot{\varepsilon} a^{-2}$. We should note that the matter (including the cold dark matter) and the radiation contributions are encoded in $\varepsilon$ due to Einstein's field equation. Therefore, the above argument could be considered as a theoretical agreement between the $\Lambda$CDM model formula (\ref {a-31}) and the equation (\ref {a-26}) which we obtained from the fundamental geometric assumptions of the Wiener fractal gravity.

\par
\section{Discussion and Review: Ricci Flow, Thermodynamics and Gravity}
\setcounter{equation}{0}

\par According to Perelman's seminal paper \cite{perelman} the Ricci flow is an entropic based differential equation. In fact, it can be seen that the Ricci flow is the gradient of Perelman's $\mathcal F$-entropy
\begin{equation} \label{F-entropy}
\mathcal F(g_{ij}(t),f(t))=\int_M \left(R + |\nabla f|^2 \right) e^{-f} d\Omega_{g(t)}
\end{equation}

\noindent via a volume preserving variation of the metric $g_{ij}(t)$.\footnote{This can be proven by employing an appropriate diffeomorphism $\varphi: M \to M$. For more discussion about the appropriate diffeomorphism see \cite{deturck} and \cite{hamilton4}.} However, from the prospects of theoretical physics, $\mathcal F$ and its first variation describes the low energy effective action in the string theory, wherein $f$ is the dilaton field.\footnote{See for example \cite{ts}, where the author shows that the generalization of Perelman's $\mathcal F$-entropy to all loop orders produces the central charge (with an extra minus sign) at the fixed points, which is in full agreement with the general claim of Zamolodchikov's c-theorem.} On the other hand, Perelman's generalized entropy functional, the so-called Perelman's $\mathcal W$-entropy
\begin{equation} \label{W-entropy}
\mathcal W(g_{ij}(t),f(t),\tau)=\frac{1}{\left( 4\pi \tau \right)^{n/2}}\int_M \left(\tau \left( R + |\nabla f|^2 \right) +f -D \right) e^{-f} d\Omega_{g(t)},
\end{equation}

\noindent as the fundamental generator of the Ricci flow has a specific interpretation as the entropy of a canonical ensemble in statistical mechanics \cite{perelman}.\footnote{Perelman's $\mathcal W$-entropy also describes the RG flow in quantum field theories. See also \cite{li-entropy, li-entropy2} for more discussions due.} 

\par On the other hand, as we explained above, the Wiener fractal measure is based on the Brownian motion which is in its own turn subject to an entropic force \cite{neu}. All in all, the emergence of the Einstein-Hilbert action in the genuine formulation of the Wiener fractal gravity as a theory of (semi-classical) gravity which combines the Hamilton-Perelman Ricci flow and the Brownian motion could be considered as the theoretical evidence for the entropic source of the gravity effects in nature. In other words, considering the Ricci flow for the evolution of the space geometry and assuming the Brownian motion of quantum states as a background framework for the dynamics of the quantum fields in the formulation of Wiener fractal gravity confirm the thermodynamical interpretation of gravity at quantum levels of nature. Moreover, we should remark that nowadays there is no doubt that classical (and semi-classical) general relativity is intimately correlated to thermodynamics. In principle, there are already enough theoretical reasons that bring up thermodynamical interpretations for Einstein's theory of gravity.\footnote{The correlation between thermodynamics and general relativity was first put forward by Bekenstein and Hawking via studying the black hole entropy in the early years of the 1970s \cite{bekenstein, hawking}. See \cite{pad, visser, jacobson} and the references therein for more recent discussions.} 

\par Thus, one may speculate that the Wiener fractal gravity has a lot of properties in common with Verlinde's entropic gravity \cite{verlinde11, verlinde16}. However, there is a profound distinction between these two approaches to gravitational force. In particular, the entropic source of the Wiener fractal gravity is due to the evolution of the space geometry via the Ricci flow, regardless of the existence and the interaction of matter fields on the space-time continuum $\mathcal M$, whereas the entropic source of Verlinde's theory is solely due to the (quantum) matter fields interactions. Actually, the latter is based on a combination of Jacobson's thermodynamical viewpoint to gravity with 't Hooft's holographic principle and interprets gravity as a consequence of the information associated with the positions of material bodies.

\par In other words, Verlinde's theory claims that gravity is not in fact a fundamental interaction, but an emergent phenomenon. Although this conclusion is confirmed by the Wiener fractal gravity, this theory secures a fundamental geometric source for the entropic root of gravity. Actually, although the space metric $g_{ij}$ (or the space-time metric $\mathbf g_{\mu \nu}$) fits with the distribution of matter fields and their interactions via the semi-classical solution of Einstein's field equation, its time evolution, which is the main reason for the emergence of the Einstein-Hilbert action in the Wiener fractal gravity has an entropic essence due to the Ricci flow. Therefore, despite their seemingly similar interpretations based on entropic origin, the Wiener fractal gravity in its current formulation is fundamentally different from Verlinde's theory of entropic gravity. 

\par However, we may ask: Is the difference between Wiener fractal gravity and Verlinde's theory really fundamental? In fact, one may enquire what is meant by the second law of thermodynamics for the geometry of space and not physical matter fields. It must be emphasized that in our formulations the second law of thermodynamics is, in fact, a geometric equation describing the law just as pointed out by Perelmean \cite{perelman} rather than a physical law for interacting matters. The distinction between Wiener fractal gravity and Verlinde's gravity stems actually from this discrepancy. However, we are not aware of the physical reason for the emergence of the geometric equation of the Ricci flow, if any, but if we could find a consistent theoretical correlation between the matter fields interactions and the Ricci flow (or Perelman's $\mathcal F$- or $\mathcal W$- entropy), then the Wiener fractal gravity could be identified with the Verlinde's theory of gravity.\footnote{Here, we must insist that the Wiener fractal gravity is inherently a non-local formulation of gravitation and this can be regarded as a fundamental distinction between our formulations and any local theory of gravity. Actually, almost the whole strategies we employed above were aimed to work out the Einstein-Hilbert action in the Wiener fractal measure as a local formulation. For instance, \textbf{Approximation I} helped us to derive a local formulation of the Jacobian terms by restricting the trace of the Ricci transformations $F_i$s in (\ref {a-6}) to the lowest eigenvalue of the Laplacian. As we will show in \cite{varshovi2}, this is the main idea to obtain the same result for Yang-Mills fields in the presence of gravity.}

\par Any way, based on the current formulation of Wiener fractal gravity, the emergence of the Einstein-Hilbert action is a mere consequence of evolving the geometric structure on $M$ and has nothing to do with the quantum field theory (i.e. the interaction terms and their low and high energy scenarios) involved in the space-time continuum $\mathcal M$. The only fundamental law that we imposed on the Wiener process of the Brownian motion of quantum states was the Ricci flow of the universe geometry. Therefore, as we emphasized above, one may conclude that the second law of thermodynamics (in its geometric formulation) is a fundamental law of nature, either more fundamental than gravitation or coming from the same basic source together with gravity, i.e, the source of the Ricci flow itself.

\par Moreover, including squared terms of time derivation, the Wiener fractal measure is essentially symmetric with respect to time-reversal transformation $t \to -t$, but, however, imposed for evolving the geometry of the background manifold $M$, the Ricci flow is, in fact, asymmetric versus reversing the time direction. Hence, Hamilton-Perelman Ricci flow is the only \emph{time's arrow-dependent} equation (i.e. an entropic source) that is included in the geometric foundations of the Wiener fractal gravity. Therefore, one may readily regard the Ricci flow of the cosmos geometry as an immediate aspect of the second law of thermodynamics. This argument and what we worked out in section III also show that gravity, despite its theoretical appearance, which is time-reversal, is inherently derived from a time-asymmetric equation.

\par The stochastic base of the Wiener fractal gravity puts forward the idea of the existence of an intimate correlation between our path-integral formulation of general relativity and the theory of stochastic gravity \cite{hu, hu2}. In fact, stochastic gravity is based on the same foundations as the Wiener fractal gravity. Actually, in stochastic gravity, the semi-classical solutions are effectively modified by including background noise from quantum fields via the Langevin equation. For instance, the familiar moving charge quantum field system in the framework of stochastic gravity leads to the so-called Abraham-Lorentz-Dirac-Langevine equation (a generalized version of the ALD equation) which describes the stochastic dynamics of the moving charge within a correct and pathology-free setting \cite{hu}.

\par Historically, stochastic (semi-classical) gravity commenced in the 1990s as the third step of a theoretical evolution of quantum gravity which its two first development stages were \emph{quantum field theory in curved spacetime} (accomplished in the 1970s) and \emph{semi-classical gravity} (established in the 1980s). Actually, the best introduction to stochastic gravity is found in \cite{hu} where the authors explain: \emph{"While semi-classical gravity is based on the semi-classical Einstein equation with the source given by the expectation value of the stress-energy tensor of quantum fields, stochastic semi-classical gravity includes also its fluctuations in a new stochastic semi-classical Einstein-Langevin equation. If the centerpiece of semi-classical gravity is the vacuum expectation value of the stress-energy tensor of a quantum field, the centerpiece in stochastic semi-classical gravity is the symmetrized stress-energy bi-tensor and its expectation value known as the noise kernel. The mathematical properties of this quantity, its physical contents in relation to the behavior of fluctuations of quantum fields in curved spacetimes, and their backreaction in the spacetime dynamics
engendering induced metric fluctuations are the main focus of this theory."}

\par Indeed, stochastic gravity could be simply regarded as a rigorous mathematical technique to work out the solution of the semi-classical Einstein's field equation that is extracted from the localized (i.e approximated) version of the genuine formula of Wiener fractal gravity. In fact, as we established above, according to the original formulation of the Wiener fractal measure with the dynamics of the space geometry due to the Ricci flow, both the quantum field fluctuations and the gravitational features are basically non-local effects, admitting no definite Lagrangian density in a genuine path-integral formulation of the Wiener fractal gravity. However, as we showed in section III, an explicit action formula for the quantum field theory on a curved space-time continuum could be extracted at the first approximation of the Wiener fractal gravity. In other words, stochastic gravity is in principle one of the best-understood frameworks to work out the solutions of the approximated version of the Wiener fractal gravity described in (\ref {aa-4}).

\par The substantial distinguishing between space and time dimensions and considering the space-time continuum as a product manifold $\mathcal M=[-T,T]\times M$ (or considering the space manifolds as the foliation leaves of $\mathcal M$), may also resemble the main ideas of Horava-Lifshitz gravity \cite{horava}. In other words, the separation of space and time dimensions, hence breaking the background local Lorentz symmetry in its general formulation is presupposed in both Horava-Lifshitz and the Wiener fractal gravity. In principle, in both theories, we witness the violation of the substantial symmetry between spacial dimensions and time at high energy levels. This symmetry-breaking is fundamental in both theories of gravity and could not be removed at all.\footnote{Even modified versions of Horava-Lifshitz gravity give rise to dynamical mechanisms for breaking the Lorentz symmetry in the UV contribution at quantum levels. See for example \cite{vag1, vag2}.} 

\par However, we should stress that this similarity only includes a part of fundamental viewpoints in the Wiener fractal gravity. In fact, there is a vast distinction between Horava's theory and the Wiener fractal gravity in their basic theoretical strategies and the mathematics of the formulation; The Wiener fractal gravity extracts the Einstein-Hilbert action as an approximate consequence of the Ricci flow, but it is fundamentally supposed in the Horava-Lifshitz gravity. In addition, the former includes non-local terms at high energy levels, but the latter is a local theory. Apart from all these differences, the Wiener fractal gravity is a semi-classical theory and does not contain quantum fluctuations of the metric components in its present form, while the Horava-Lifshitz gravity is essentially a theory for quantum gravity.

\par Actually, Horava's theory respects some significant admitted properties for quantum field theory such as the necessity for the Lagrangian density to only consist of local terms and to maintain the renormalizability of the theory by power-counting at the Lifshitz point $z=3$, but the Wiener fractal gravity is a highly non-local theory at high energy levels which does not even admit a definite form of the Lagrangian density in its original expression. Although both try to achieve a consistent theory of (quantum) gravity within the path-integral formulation and also both address the breaking of the (local) Lorentz symmetry and general diffeomorphism invariance of the theory, Horava's strategy to overcome the renormalizability problem is somehow active. He breaks the Lorentz invariance (and diffeomorphism invariance) by hand, puts forward the fixed Lifshitz point $z=3$ according to some evidence coming from condensed matter, and introduces and incorporates some foliation-preserving diffeomorphism invariant terms for bringing gravity into the quantum field theory in some renormalizable manner.

\par On the other hand, in Wiener fractal gravity we behave as passive. We do not impose any preferred geometric structure (such as some definite symmetries) and theoretical/geometric constraints into the theory except the Ricci flow. We also assume that the quantum field propagation along time is fundamentally a stochastic process that is partly subject to the fractal structure of the field and partly to some dominating entropic force due to the Ricci flow. Therefore, the most geometric structures and theoretical constraints of the theory emerge naturally from the essential properties of the Wiener Brownian process and the entropic effect of the Ricci flow. In fact, the theory basically respects Heisenberg's uncertainty principle of quantum mechanics (due to its intrinsic stochastic property) and is fundamentally entropic (i.e. breaks the time-reversal symmetry). 

\par Furthermore, in the Wiener fractal gravity, we surprisingly showed that the source of gravity is intrinsically different from that of the three fundamental forces of nature which have been already formulated in the Standard Model, i.e. the electromagnetic and the weak and the strong nuclear forces. Thus, the viewpoint of the Wiener fractal gravity to quantum interactions of nature may resolve or interpret the question of: \emph{"Why we cannot unify gravity with other fundamental forces of nature?"} Indeed, based on the current formulation of Wiener fractal gravity and upon what we have done above the answer is: \emph{Because the quantum fundamental forces of nature stem from Brownian propagation of fractal quantum fields along time, whereas gravity is basically rooted in entropic evolution of the geometry of the universe}. Obviously, they address two fiercely distinct scales of nature, as we expected before.

\par
\section{Summary and Conclusion}
\setcounter{equation}{0}

\par In this paper, we established a well-defined path-integral formulation for quantum field theory in the presence of gravity by employing three basic theoretical/mathematical mechanisms: \\

\par \textbf{a)} \emph{The asymptotic properties of Fourier-Laplace coefficients of the Weierstrass-like fractal functions via the fractal norm on closed Riemannian manifolds.}
\par \textbf{b)} \emph{The Wiener stochastic process for Brownian motion of quantum states on a closed Riemannian manifold by employing the Wiener fractal measure.}
\par \textbf{c)} \emph{The entropic-based Hamilton-Perelman Ricci flow of the Riemannian metric on a closed Riemannian manifold.} \\

\noindent We proved that the stochastic process of Brownian motion of quantum states on a closed Riemannian manifold with dynamical geometry due to the Ricci flow causes the emergence of the Einstein-Hilbert action in the path-integral formulation of the corresponding quantum field theory on the curved space-time continuum within a generalized well-defined Wiener probability measure. The resulting theory, the so-called \emph{Wiener fractal gravity}, could be regarded as a fundamental formulation of quantum field theory in the presence of semi-classical gravity. We also established that the emergence of the Einstein-Hilbert Lagrangian density has nothing to do with the background interactions of the quantum field theory and has a merely entropic source due to the Ricci flow.

\par We discussed the different sources of gravity and the other fundamental forces of nature based on the mentioned stochastic framework on dynamical Riemannian manifolds. Also, by computing the coupling constant of the Einstein-Hilbert action we interpreted the hierarchy problem in terms of the background geometric structure of the Ricci flow. Then, we worked out the cosmological constant according to Hamilton's theorem for 3-manifolds and showed that in $D=3$ dimensions this constant is positive and extremely small. We also argued that in the Wiener fractal gravity the cosmological constant is a pure entropic/geometric term and is allegedly independent of the zero-point energy of the involved quantum field theories.

\par The correlation of the solutions of the corresponding Ricci flow and the achievements of the $\Lambda$CDM Model was discussed and we found some specific similarities to the FLRW metric and its cosmological features. Hubble's law and the acceleration of the universe expansion were worked out due to the Ricci flow and it was argued that the dark energy (hence the inflation) has an entropic source. Afterward, we compared our achievements with three important theories of gravity; \textbf{1)} Verlinde's entropic gravity, \textbf{2)} Stochastic gravity, and \textbf{3)} Horava-Lifshitz gravity. We found both basic similarities and fundamental differences between these theories and the Wiener fractal gravity.

\par Based upon the above results, we are optimistic that the above formulation could pave the way to produce a well-defined framework of quantum gravity. Indeed, within the present form of the Wiener fractal gravity, the gravitational fields have no quantum (fractal) fluctuations, and their physical effects in Einstein's field equation are regarded in terms of the expectation values of the involved quantum fields via a semi-classical viewpoint. Actually, to obtain a well-defined theory of quantum gravity within the framework of Wiener fractal gravity one should first consider the Brownian motion of the metric via some fractal norm for positive-definite symmetric $(0,2)$-tensor fields on $M$. Then, one has to localize the non-local fluctuations of the metric appearing in the formulation. This strategy, as we have shown in \cite{varshovi2}, will give rise to the appearance of some specific numbers of Grassmannian spinor fields, hence causing the emergence of fermionic Majorana quantum fields accompanying the bosonic gravitational fields within the genuine formulation of the Wiener path-integral. The approach faces a number of theoretical problems due to the fractality of the gravitational fields and the corresponding Laplacian operator but it could be regarded as a possible procedure to produce a path-integral formulation of supergravity in terms of a Wiener fractal probability measure.






\begin{thebibliography}{99}
\bibitem{azami} S. Azami, \emph{Variation of the First Eigenvalue of $(p, q)$-Laplacian along the Ricci-Harmonic Flow}, Int. J. Nonlin. Anal. Appl. 12, No. 2: 193-204, 2021.
\bibitem{bekenstein} A. Bekenstein, \emph{Black Holes and the Second Law}, Lett. Nuovo Cimento. 4 (15): 99-104, 1972.
\bibitem{berger} M. Berger, \emph{Sur le Spectre $\acute{d}$une Vari$\acute{e}$t$\acute{e}$ Riemannienne}. C.R. Acad. Sci. Paris 263: 13-16, 1963.
\bibitem{berger2} M. Berger, \emph{Sur les Premieres Valeurs Propres des Vari$\acute{e}$t$\acute{e}$s Riemannienes}, Compositio Math. 26 (2): 129-149, 1973.
\bibitem{berger et al} M. Berger, P. Gauduchon, and E. Mazet, \emph{Le Spectre $\acute{d}$une Vari$\acute{e}$t$\acute{e}$ Riemannienne}, Springer, 1971. 
\bibitem{besse} A. L. Besse, \emph{Einstein Manifolds}, Springer, 2008.
\bibitem{brans-dicke} C. H. Brans, and R. H. Dicke, \emph{Mach's Principle and a Relativistic Theory of Gravitation}, Phys. Rev. 124 (3): 925-935, 1961.
\bibitem{buser} P. Buser, \emph{A Note on the Isoperimetric Constant}, Ann. Sci. $\acute{E}$cole Norm. Sup. 15, No. 2: 213-230, 1982.
\bibitem{cai} Y. F. Cai, J. Liu, and H. Li, \emph{Entropic Cosmology: A Unified Model of Inflation and Late-Time Acceleration}, Phys. Lett. B. 690 (3): 213–219, 2010.
\bibitem{cao} X. D. Cao, \emph{First Eigenvalues Geometric Operators under the Ricci Flow}, Proc. Amer. Math. Soc. 136: 4075-4078, 2008.
\bibitem{vag2} A. Casalino, M. Rinaldi, L. Sebastiani, S. Vagnozzi, \emph{Alive and Well: Mimetic Gravity and a Higher-Order Extension in Light of GW170817}, Class. Quant. Grav. 36: 017001, 2019.
\bibitem{chang} Z. Chang, M. H. Li, X. Li, \emph{Unification of Dark Matter and Dark Energy in a Modified Entropic Force Model}, Commun. Theor. Phys. 56 (1): 184-192, 2011. 
\bibitem{chavel} I. Chavel, \emph{Eigenvalues in Riemannian Geometry}, Academic Press, 1984.
\bibitem{cheeger} J. Cheeger, \emph{A Lower Bound for the Smallest Eigenvalue of the Laplacian}, In: Problems in Analysis (Papers dedicated to Salomon Bochner, 1969), Princeton University Press, Princeton, N. J., pp. 195–199, 1970.
\bibitem{vag1} G. Cognola, R. Myrzakulov, L. Sebastiani, S. Vagnozzi, S. Zerbini, \emph{Covariant Horava-Like and Mimetic Horndeski Gravity: Cosmological Solutions and Perturbations}, Class. Quant. Grav. 33: 225014, 2016.
\bibitem{dem} M. Demianski, N. Sanchez, and Y. N. Parijskij, \emph{Topology of the Universe and the Cosmic Microwave Background Radiation}, Proceeding of the NATO Advanced Study Institute 130, Springer, 2003.
\bibitem{deshmukh} S. Deshmukh and A. Al-Eid, \emph{Curvature Bounds for the Spectrum of a Compact Riemannian manifold of Constant Scalar Curvature}, J. Geom. Anal. 15(4): 589-606, 2005.
\bibitem {deshmukh0} S. Deshmukh, \emph{Eigenvalues of the Laplacian Operator on Compact Riemannian Manifolds}, Arab. J. Math. Sc. Vol. 13: 39-65, 2007.
\bibitem{deturck} D. DeTurck, \emph{Deforming Metrics in the Direction of their Ricci Tensors}, In: H. D. Cao, B. Chow, S. C. Chu and S. T. Yau (Eds.), \emph{Collected Papers on Ricci Flow}, Series in Geometry and Topology, 37, International Press, 2003.
\bibitem{dewitt} B. S. DeWitt, \emph{Quantum Field Theory in Curved Spacetime}, Phys. Rep. 19, No. 6: 295-357, 1975.
\bibitem{dicerbo} L. F. Di Cerbo, \emph{Eigenvalues of the Laplacian under the Ricci Flow}, Rend. Mat. VII, Vol. 27, Rome, 2007.
\bibitem{donnelly} H. Donnelly, \emph{Eigenfunctions of the Laplacian on Compact Riemannian Manifolds}, Asian J. Math. Vol. 10, No. 1: 115–126, 2006.
\bibitem{dowker} F. Dowker, \emph{Topology Change in Quantum Gravity}, 2002 [arXiv:gr-qc/0206020].
\bibitem{dz} V. Dzhunushaliev, \emph{Quantum Wormhole as a Ricci Flow}, Int. J. Geom. Meth. Mod. Phys. Vol. 06, No. 06: 1033-1046, 2009.
\bibitem{dz2} V. Dzhunushaliev, V. Folomeev, \emph{Masking Singularities in Weyl Gravity and Ricci Flows}, Eur. J. Phys. C 81: 387-, 2021.
\bibitem{dz3} V. Dzhunushaliev, N. Serikbayev, and R. Myrzakulov, \emph{Topology Change in Quantum Gravity and Ricci Flows}, 2010 [	arXiv:0912.5326 [gr-qc]].
\bibitem{egidi} M. Egidi, S. Liu, F. M$\ddot{u}$unch, and N. Peyerimhoff, \emph{Ricci Curvature and Eigenvalue Estimates for the Magnetic Laplacian on Manifolds}, Commun. Anal. Geom. Vol. 29, No. 5, 2021.
\bibitem{elsoufi} A. El Soufi, and S. Ilias, \emph{Laplacian Eigenvalue Functionals and Metric Deformations on Compact Manifolds}, J. Geom. Phys. 58: 89-104, 2008.
\bibitem{fang} S. Fang, F. Yang, and P. Zhu, \emph{Eigenvalues of Geometric Operators Related to the Witten Laplacian under the Ricci Flow}, Glas. Math. J.: 1-9, 2017.
\bibitem{folland} G, B. Folland, \emph{Real Analysis; Modern Techniques and Their Applications}, 2nd Ed., John Wiley and Sons, 1999.
\bibitem{top1} A. Frenkel, P. Horava, and S. Randall, \emph{Topological Quantum Gravity of the Ricci Flow}, preprint, 2020 [arXiv:2010.15369 [hep-th]].
\bibitem{top2} A. Frenkel, P. Horava, and S. Randall, \emph{The Geometry of Time in Topological Quantum Gravity of the Ricci Flow}, preprint, 2020 [arXiv:2011.06230 [hep-th]].
\bibitem{top3} A. Frenkel, P. Horava, and S. Randall, \emph{Perelman's Ricci Flow in Topological Quantum Gravity}, preprint, 2020 [arXiv:2011.11914 [hep-th]].
\bibitem{hier} A. Fowlie, C. Balazs, G. White, L. Marzola, and M. Raidal, \emph{Naturalness of the Relaxion Mechanism}, JHEP 2016 (8): 100, 2016.
\bibitem{fri} J. A. Frieman, M. S. Turner, and D. Huterer, Dragan, \emph{Dark Energy and the Accelerating Universe}, Ann. Rev. Astr. Astrophys. 46 (1): 385-432, 2008.
\bibitem{gallot} S. Gallot, D. Hulin, and J. Lafontaine, \emph{Riemannian Geometry}, 2nd Ed., Springer-Verlag, 1990.
\bibitem{gilkey1} P. B. Gilkey, \emph{Curvature and the Eigenvalues of the Laplacian for Elliptic Complexes}, Adv. Math. 10: 344-382, 1973. 
\bibitem{gilkey2} P. B. Gilkey, \emph{The Spectral Geometry of Riemannian Manifolds}, J. Diff. Geom. 10: 601-618, 1975.
\bibitem{gilkey3} P. B. Gilkey, \emph{Recursion Relations and the Asymptotic Behavior of the Eigenvalues of the Laplacian}, Comp. Math. Vol. 38, Fasc. 2: 201-240, 1979.
\bibitem{sobolev} L. Giovanni, \emph{A First Course in Sobolev Spaces}, Graduate Studies in Mathematics. 105. American Mathematical Society, 2009.
\bibitem{yau} A. Grigor'yan, Y. Netrusov, and S. T. Yau, \emph{Eigenvalues of Elliptic Operators and Geometric Applications}, Surv. Diff. Geom. IX: 147-217, 2004.
\bibitem{graf} W. Graf, \emph{Ricci Flow Gravity}, PMCPhys. A 1: 3, 2007.
\bibitem{hamilton1} R. S. Hamilton, \emph{Three-Manifolds with Positive Ricci Curvature}, J. Diff. Geom. 17: 255-306, 1982.
\bibitem{hamilton2} R. S. Hamilton, \emph{Four-Manifolds with Positive Curvature Operator}, J. Diff. Geom. 24: 153-179, 1986.
\bibitem{hamilton3} R. S. Hamilton, \emph{The Ricci Flow on Surfaces}, Contemp. Math. 71, Amer. Math. Soc., Providence, RI: 237-262, 1988.
\bibitem{hamilton4} R. S. Hamilton, \emph{Formation of Singularities in the Ricci Flow}, Surv. Diff. Geom. II: 7-136, 1995.
\bibitem{hardi} G. H. Hardy, \emph{Weierstrass’s Non-Differentiable Function}, Trans. Amer. Math. Soc. 17, 3: 301-325, 1916.
\bibitem{hawking} S. Hawking, \emph{Particle Creation by Black Holes}, Commun. Math. Phys. 43 (3): 199-220, 1975.
\bibitem{henrot} A. Henrot, \emph{Minimization Problems for Eigenvalues of the Laplacian}, J. Evol. Eqs. 3: 443-461, 2003.
\bibitem{hobson} M. P. Hobson, G. P. Efstathiou, and A. N. Lesenby, \emph{General Relativity: An Introduction for Physicists}, Cambridge University Press, 2006.
\bibitem{holland} S. Hollands, and R. M. Wald, \emph{Quantum Fields in Curved Spacetime}, In: A. Ashtekar, B. Berger, J. Isenberg, and M. MacCallum (Eds.), \emph{General Relativity and Gravitation: A Centennial Perspective}, pp. 513-552, Cambridge: Cambridge University Press, 2015.
\bibitem{horava} P. Horava, \emph{Quantum Gravity at a Lifshitz Point}, Phys. Rev. D. 79 (8): 084008, 2009.
\bibitem{hou} S. Hou, and S. Yang, \emph{Eigenvalues of the Laplace Operator with Potential under the Backward Ricci Flow on Locally Homogeneous 3-Manifolds}, Manus. Math. 2021.
\bibitem{hu2} B. L. Hu, and E. Verdaguer, \emph{Stochastic Gravity: Theory and Applications}, Liv. Rev. Relativ. 7(1): 3, 2004.
\bibitem{hu} B. L. Hu, and E. Verdaguer, \emph{Semiclassical and Stochastic Gravity: Quantum Field Effects on Curved Spacetime}, Cambridge University Press, 2020.
\bibitem{hubble} E. Hubble, \emph{A Relation between Distance and Radial Velocity among Extra-Galactic Nebulae}, Proc. Natl. Acad. Sci. USA 15(3):168–173, 1929.
\bibitem{huisken} G. Huisken, \emph{Ricci Deformation of the Metric of a Riemannian Manifold}, J. Diff. Geom. 21: 47-62, 1985.
\bibitem{hwang} S. Hwang, J. Chang, and G. Yun, \emph{Variational Characterizations of the Total Scalar Curvature and Eigenvalues of the Laplacian}, Pac. J. Math. 261, No. 2: 395-415, 2013.
\bibitem{isi} J. M. Isidro, J. L. G. Santander, and P. Fernandez de Cordoba, \emph{Ricci Flow, Quantum Mechanics and Gravity}, Geom. Meth. Mod. Phys. Vol. 06, No. 03: 505-512, 2009.
\bibitem{jacobson} T. Jacobson, \emph{Thermodynamics of Spacetime: The Einstein Equation of State}, Phys. Rev. Lett. 75: 1260-1263, 1995.
\bibitem{lang} C. Lange, S. Liu, O. Post, and N. Peyerimhoff, \emph{Frustration Index and Cheeger inequalities for Discrete and Continuous Magnetic Laplacians}, Calc. Var. Par. Diff. Eq. 54, No. 4: 4165-4196, 2015.
\bibitem{ledoux} M. Ledoux, \emph{Spectral Gap, Logarithmic Sobolev Constant, and Geometric Bounds}, Surv. Diff. Geom. IX: 219-240, 2004.
\bibitem{levin} J. Levin, \emph{Topology and the Cosmic Microwave Background}, Phys. Rept. 365: 251-333, 2002.
\bibitem{li-entropy2} X. D. Li, \emph{Perelman’s Entropy Formula for the Witten Laplacian on Riemannian Manifolds via Bakry–Emery Ricci Curvature}, Math. Ann. 353:403–437, 2012.
\bibitem{li-entropy} X. D. Li, \emph{From the Boltzmann H-Theorem to Perelman’s W-Entropy Formula for the Ricci Flow}, Emer. Top. Diff. Equ. App.: 68-84, 2013.
\bibitem{lic} A. Lichnerowicz, \emph{G$\acute{e}$om$\acute{e}$trie des Groupes de Transformations}, Trav. Rech. Math. 3, 1958.
\bibitem{li-yau} P. Li, and S. T. Yau, \emph{Eigenvalues of a Compact Riemannian Manifold}, AMS Proc. Symp. Pure Math. 36: 205-239, 1980. 
\bibitem{lulli} M. Lulli, A. Marciano, and X. Shan, \emph{Stochastic Quantization of General Relativity a la Ricci-Flow}, preprint, 2021 [arXiv:2112.01490 [gr-qc]].
\bibitem{singer} H. P. McKean, and I. M. Singer, \emph{Curvature and the Eigenforms of the Laplacian}, J. Diff. Geo. 1: 43-69, 1967.
\bibitem{mccann} R. J. McCann, and P. M. Toping, \emph{Ricci Flow, Entropy and Optimal Transportation}, Amer. J. Math. 132, No. 3: 711-130, 2010.
\bibitem{morison} C. Morison, \emph{Uniformisation as a Bridge Between Ricci Flow and General Relativity in Two Spatial Dimensions}, PhD. Thesis, University of Amsterdam, 2020.
\bibitem{neu} R. M. Neumann, \emph{Entropic Approach to Brownian Movement}, Amer. J.
Phys. 48 (5): 354-357, 1980.
\bibitem{pad} T. Padmanabhan, \emph{Thermodynamical Aspects of Gravity: New Insights}, Rep. Prog. Phys. 73 (4): 6901, 2010.
\bibitem{peebles} P. J. E. Peebles, \emph{Cosmology’s Century; An Inside History of Our Modern Understanding of the Universe}, Princeton University Press, 2020.
\bibitem{perelman} G. Perelman, \emph{The Entropy Formula for the Ricci Flow and its Geometric Applications}, preprint, 2002.
\bibitem{poor} W. A. Poor, \emph{Differential Geometric Structures}, Dover Publications, 1981.
\bibitem{simon} U. Simon, \emph{Curvature Bounds for the Spectrum of a Closed Einstein Spaces}, Can. J. Math. 4: 1087-1091, 1978. 
\bibitem{ts} A.A. Tseytlin, \emph{On Sigma Model, RG flow, Central Charge Action and Perelman's Entropy}, Phys. Rev. D 75: 064024, 2007.
\bibitem{varshovi1} A. A. Varshovi, \emph{Brownian Motion in the Hilbert Space of Quantum States and the Stochastically Emergent Lorentz Symmetry: A Fractal Geometric Approach from Wiener Process to Formulating Feynman's Path-Integral Measure for Relativistic Quantum Fields}, Int. J, Geom. Meth. Mod. Phys., 2024, DOI: 10.1142/S0219887824502025
\bibitem{varshovi2} A. A. Varshovi, \emph{Brownian Motion in the Hilbert Space of Gauge Field Quantum States and Stochastically Emergent Yang-Mills Action with Dirac Matter Fields: Formulating a Well-Defined Feynman's Path-Integral Measure for the Standard Model in the Presence of Gravity}, preprint.
\bibitem{verlinde11} E. P. Verlinde, \emph{On the Origin of Gravity and the Laws of Newton}, JHEP (4): 29, 2011.
\bibitem{verlinde16} E. P. Verlinde, \emph{Emergent Gravity and the Dark Universe}, SciPost Phys. 2, 016, 2017.
\bibitem{visser} M. Visser, \emph{Conservative Entropic Forces}, JHEP
(10): 140, 2011.
\bibitem{wald1} R. M. Wald, \emph{The Backreaction Effect in Particle Creation in Curved Spacetime}, Commun. Math. Phys. 54:1-19, 1977.
\bibitem{wald2} R. M. Wald, \emph{Trace Anomaly of a Conformally Invariant Quantum Field in Curved Space-Time}, Phys. Rev. D. 17:1477-1484, 1978.
\bibitem{warner} F. W. Warner, \emph{Foundations of Differentiable Manifolds and Lie Group}, Springer-Verlag, 1983.
\bibitem{weinberg} S. Weinberg, \emph{Gravitation and Cosmology: Principles and Applications of the General Theory of Relativity}, John Wiley and Sons, 1972.
\bibitem{weinberg-cos} S. Weinberg, \emph{The Cosmological Constant Problem}, Rev. Mod. Phys. 61, No. 1: 1-23, 1989.
\bibitem{weierstrass} K. Weierstrass, \emph{$\ddot{U}$ber Continuirliche Functionen eines Reellen Arguments, die f$\ddot u$r Keinen Werth des Letzeren einen Bestimmten Differentialquotienten Besitzen}, Mathematische Werke von Karl Weierstrass, 2, Berlin, Germany: Mayer and Mller, pp. 71-74, 1895, English translation: G. Edgar, \emph{On Continuous Functions of a Real Argument That Do not Possess a Well-Defined Derivative for Any Value of Their Argument, Classics on Fractals}, Studies in Nonlinearity, Addison-Wesley Publishing Company, pp. 3-9, 1993.
\bibitem{weyl} H. Weyl, \emph{$\ddot{U}$ber die Asymptotische Verteilung der Eigenwerte}, Nachrichten der K$\ddot{o}$niglichen Gesellschaft der Wissenschaften zu Göttingen, pp. 110-117, 1911.
\bibitem{weyl2} H. Weyl, \emph{The Classical Groups}, Princeton University Press, 1946.
\bibitem{wiener} N. Wiener, \emph{Norbert Wiener: Collected Works, Volume I; Mathematical Philosophy and Foundations; Potential Theory; Brownian Movement, Wiener Integrals, Ergodic and Chaos Theories, Turbulence and Statistical Mechanics}, P. Masani (Ed.), MIT Press, 1976
\bibitem{woolgar} E. Woolgar, \emph{Some Applications of Ricci Flow in Physics}, Can. J. Phys. 86: 645, 2008.
\bibitem{yau0} S.T. Yau, \emph{Isoperimetric Constants and the First Eigenvalue of a Compact 
Riemannian Manifold}, Ann. Scient. Ec. Norm. Sup. 4: 487-507, 1985. 
\bibitem{yoo} M. Y. Yoo, \emph{Unexpected Connections}, Eng. Sc. LXXIV1: 30, 2011.


\end{thebibliography}
\end{document}